\pdfoutput=1
\documentclass[noinfoline,stslayout]{imsart}
\usepackage{amsmath}
\usepackage{graphicx}
\usepackage{amssymb}
\usepackage{natbib}
\usepackage{algorithm}
\usepackage{subfigure}
\usepackage{algpseudocode}
\usepackage{bm}
\usepackage{url}

\DeclareMathOperator{\tr}{\tau}

\begin{document}

\begin{frontmatter}

\title{Accelerating inference for diffusions observed with measurement error and large sample sizes using Approximate Bayesian Computation}

\begin{aug}
 \author{\snm{{\sc Umberto Picchini}}}
 \affiliation{Centre for Mathematical Sciences, Lund University, Sweden}
 \author{\snm{{\sc Julie Lyng Forman}}}
 \affiliation{Department of Biostatistics, University of Copenhagen, Denmark}
\end{aug}

\begin{abstract}
In recent years dynamical modelling has been provided with a range
of breakthrough methods to perform exact Bayesian inference. However it is often computationally unfeasible to apply exact statistical
methodologies in the context of large datasets and complex
models. This paper considers a nonlinear
stochastic differential equation model observed with correlated measurement errors and an
application to protein folding modelling.
An Approximate Bayesian Computation (ABC) MCMC algorithm is suggested to
allow inference for model parameters within reasonable time constraints. The ABC algorithm uses
simulations of ``subsamples'' from the assumed data generating model as well as a so-called ``early rejection'' strategy to
speed up computations in the ABC-MCMC sampler. Using a considerate amount of subsamples does not seem to degrade the quality of the inferential results for the considered applications. A simulation
study is conducted to compare our strategy with exact Bayesian inference, the latter resulting two orders of magnitude slower than ABC-MCMC for the considered setup.
Finally the ABC algorithm is applied to a large size protein data. 
The suggested methodology is fairly general and not limited to the
exemplified model and data.
%models expressed as sums of two diffusions.
\end{abstract}

\end{frontmatter}


\section{Introduction}

In the so-called ``Big Data'' era we face the need and the opportunity to extract information provided by a steadily increasing amount of data, as produced by e.g. \textit{in-silico} and \textit{in-vivo} experiments, to describe real-world systems at previously unattainable resolutions. As the size of datasets requiring analysis increases, so must the statistical techniques used to analyse them be able to efficiently handle the increase in scale. Standard statistical approaches, both classical and Bayesian, were not designed with this in mind and statisticians now have to consider models of adequate complexity while trying to obtain inferential results within reasonable time limits.

In recent years statistical inference for dynamical modelling has been provided with powerful tools to perform exact inference on models of considerable complexity, thanks to sequential Monte Carlo methods embedded within Markov chain Monte Carlo (MCMC) algorithms \cite{andrieu2010particle} as well as ``likelihood-free'' methods \cite{breto2009time,golightly2011bayesian}, see section \ref{sec:issues-exact-bayesian} for more details. Such methods have flourished in the Bayesian community and have pushed the exploration for possibilities previously unrealistic to contemplate.
However these computational methods usually don't scale well enough to match the increasing sizes of datasets. In this work we exemplify inference for a stochastic dynamical model describing protein dynamics time series data approximately of size 25,000, and even if such size is not large enough to be considered a typical example of ``Big Data'', it has been a challenge for us to perform inference for a particular nonlinear stochastic differential equation (SDE) model observed with correlated measurement error. The use of exact methods in our application was not feasible, without reverting to a rather arbitrary subsample of the available data. Similar difficulties are expected in applications in systems biology and bioinformatics. 

Here we present a strategy to rely on the full data-set without having to simulate trajectories for the latent process of the same size as the data. The considered inferential framework is approximate Bayesian computation (ABC) within an MCMC algorithm, where acceptance of simulated trajectories and corresponding generating parameters is regulated by the use of specific ``summary statistics''. When the chosen summary statistics applied on (relatively short) simulated trajectories approximately match the summary statistics for the (much larger) observed dataset, the proposed parameter has a higher probability to be accepted. This mechanism thus enable approximate inference for arbitrarily large datasets, as the summary statistics for the real data need to be computed only once, whereas during the ABC-MCMC algorithm statistics for simulated datasets are relatively cheap to compute, due to the shorter size of the artificial trajectories. An analysis of protein folding data is presented, based on a recent model expressed as a sum of two diffusion processes \cite{fs:14}, hereafter denoted ``diffusion observed with measurement error''. Inference via ABC is performed on such data. A simulation study for a smaller dataset is also performed, comparing ABC against exact inference obtained via particle MCMC methods \citep{andrieu2010particle}.

\section{Diffusion observed with measurement error}\label{sec:sum-of-diffusions}
As an example of a fairly complex dynamical model, we consider a
nonlinear diffusion model observed with measurement error. The model
was introduced by \cite{fs:14} to model the dynamics of a particular
protein folding problem which is further investigated in section
\ref{sec:real-application}. The stationary distribution of the
nonlinear diffusion is bimodal in order to reflect
the two regimes of the protein, \textit{folded} and \textit{unfolded}.
To be specific, let the {\em observable stochastic process} $\{Z_t\}$ be defined by
\begin{equation}
\begin{cases}
Z_t = \tau_{\psi}(X_t) + U_t, \qquad t\geq t_0 \\
dU_t = -\kappa U_tdt + \sqrt{2\kappa\gamma^2}dW_t, \qquad U_{t_0}=0 \\
dX_t = -\theta X_tdt + \sqrt{2\theta}dB_t, \qquad X_{t_0}=x_{t_0}\\
\end{cases}
\label{eq:state-space}
\end{equation}
where the {\em error process} $\{U_t\}$ is a Ornstein-Uhlenbeck
(OU) process with stationary mean zero, stationary variance $\gamma^2$ and autocorrelation function
$\rho_U(t)=e^{-\kappa t}$, the {\em latent process} $\{X_t\}$ is yet another
OU process with stationary mean zero, unit variance and autocorrelation function
$\rho_X(t)=e^{-\theta t}$, $\{W_t\}$ and $\{B_t\}$ are independent
Brownian motions. The transformation $\tau_{\psi}(\cdot)$ with
${\bm \psi}=(\alpha,\mu_1,\mu_2,\sigma_1,\sigma_2)$ is given by
$\tau_{\psi}(x) = (F_{\psi}^{-1}\circ \Phi)(x)$ where
\begin{equation}
F_{\psi}(y) = \alpha\cdot\Phi\left(\frac{y-\mu_1}{\sigma_1}\right)
     + (1-\alpha)\cdot\Phi\left(\frac{y-\mu_2}{\sigma_2}\right) \label{eq:mixture-model}
\end{equation}
and $\Phi(\cdot)$ denotes the cumulative distribution
function of the standard normal distribution. Note that the
transformation $\tr_{\psi}(\cdot)$ maps the invariant
$N(0,1)$-distribution of the OU-process $\{X_t\}$ to a bimodal mixture
of normal distributions with modes at $\mu_1$ and $\mu_2$ and mixture
parameter $\alpha\in (0,1)$. In other words, $\tau_\psi(X_t)$ is the
$\Phi(X_t)$-percentile of the two-components Gaussian mixture having
cumulative distribution function \eqref{eq:mixture-model}.  
It is important to notice that the model has a simple latent structure
arising from the fact that both the error process $\{U_t\}$ and the nonlinear diffusion $\{X_t\}$ are
OU processes, where one has been transformed to match the desired stationary distribution of the data.
Recall that an OU process has Gaussian transition densities, for example for $\{X_t\}$ we have
\begin{equation}
\label{eq:ou-trans}
X_t|X_s=x \sim N(x\cdot e^{-\theta\Delta_t},1-e^{-2\theta\Delta_t}),
\quad \textnormal{ with }\Delta_t=t-s
\end{equation}
for $s<t$. Further note that the process $\{Z_t\}_{t\geq 0}$ is able to display {\em multi-scale}
behaviour. Whenever $0<\theta\ll\kappa$, the error process $\{U_t\}$
dominates the dynamics of the observable process on the short time
scale, while the  latent nonlinear diffusion $\{\tau_\psi(X_t)\}$
determines the observed behaviour on the long time scale. We refer to
\cite{PavStuart,Azencott} for further discussion of
multi-scale models and the difficulties related to performing
statistical inference.

Please note that the statistical methodology discussed in this paper
applies to a much wider range of processes than the exemplified model
\eqref{eq:state-space}. In particular, the transformation could be
replaced by one targeting other distributions than the bimodal normal
mixture or the process $\{\tau_\psi(X_t)\}$ could be replaced by an
entirely different diffusion, e.g.\ a double-well potential model or
a nonlinear diffusion model as considered by \cite{yas:96}. More
general partially observed and multi-scale diffusions such as the ones
presented in \cite{stuart,Eijnden} could also be
considered. The motivation for choosing model \eqref{eq:state-space}
is due to the fact that it yields explicit formulae for the mean
passage times, see \cite{fs:14}, which are important for estimating
the folding and unfolding rates of the protein data (section
\ref{sec:real-application}). From the perspective of the protein
folding problem, the model has the further advantage that the
nonlinear latent diffusion displays increased volatility inbetween the
modes which is in accordance with the empirical finding of
state-dependent diffusion in protein reaction coordinates, see
\cite{bh:10}. Finally, \cite{fs:14} found that the diffusion with
error model was able to fit the protein data satisfactory both on the
short and the long time scale, which was not the case with any plain
diffusion model.

\section{Issues with exact Bayesian inference}\label{sec:issues-exact-bayesian}
Consider the problem of making inference for the parameter
${\bm \eta}=(\theta,\kappa,\gamma,{\bm \psi})$ of the nonlinear diffusion with
error model described in Section \ref{sec:sum-of-diffusions}. Denote with ${\bm z}=\{z_{0},z_{1},\ldots,z_{n}\}$ a set of discrete observations from $\{Z_t\}$ and with ${\bm x}=\{x_{0},x_{1},\ldots,x_{n}\}$ corresponding unobserved values from $\{X_t\}$. 
The likelihood function of ${\bm \eta}$ based on ${\bm z}$ is
\begin{align}
  \label{eq:full-likelihood}
L({\bm \eta})&=
  p({\bm z}|{\bm \eta})=\prod_{i=1}^np(z_{i}|z_{0},z_1,...,z_{i-1};{\bm \eta})\\
  &=\int p(z_0,...,z_n|\tau(x_0),...,\tau(x_n);{\bm \eta})p(\tau(x_0),...,\tau(x_n)|{\bm \eta})d\tau(x_0)\cdots d\tau(x_n)\nonumber\\
 &= \int p(z_0,...,z_n|\tau(x_0),...,\tau(x_n))\prod_{i=1}^n p(\tau(x_i)|\tau(x_{i-1});{\bm \eta}) d\tau(x_0)\cdots d\tau(x_n)\nonumber
\end{align}
where the product in the last integrand is due to the Markov property of $\{X_t\}$.
%($\{X_t\}$ is Markov hence so is $\{\tau(X_t)\}$ because $\tau$ is invertible).
This likelihood function is neither explicitly known nor easy to
approximate. For this reason we wish to consider a Bayesian approach
for doing inference on ${\bm \eta}$.
Unfortunately, as discussed below, several difficulties related to our
specific application prevent using conventional exact methodology.
Firstly, due to the autocorrelation in $\{U_t\}$, the observations
${\bm z}$ are \textit{not} conditionally independent given the latent state
${\bm x}$. This obstructs the use of most methods available for
state-space models (aka Hidden Markov
Models). In \cite{andrieu2010particle} it has been shown how to use sequential
Monte Carlo (SMC) methods for a class of models larger than
state-space models by use of the particle MCMC methodology.  In
principle, particle MCMC algorithms plug
an SMC approximation to \eqref{eq:full-likelihood} into an MCMC procedure for inference on ${\bm \eta}$, state variables or both. When such approximation is an unbiased estimate of the likelihood we are rewarded with
exact Bayesian inference, regardless the number of particles used in the SMC step. However, since in our case $n\approx 2.5
\times 10^4$, this approach is not practically feasible as it would
take several weeks of computation on our hardware, depending on the number of particles
used. To be specific, we
initially implemented particle MCMC approach with the
adaptation suggested in \cite{golightly2011bayesian}, suitable for
Bayesian inference for diffusion models. Even when using only 10
particles and writing our program in the \textsc{Julia} language
\citep{julia} (in some cases comparable to C++ in terms of
performance), the result was far too slow to be worthwhile. It has to be
noted, though, that we have not exploited available GPUs
implementations such as \cite{murray2013bayesian}, which are likely to
reduce the computational cost.

Without reverting to SMC methods, a class of methods that often gives
satisfactory results is the one enabling so-called ``likelihood-free'' inference, see section 9.6 in \cite{d.wilkinson(2012)}. This
is sometimes referred to as ``plug-and-play''
\cite{ionides2006inference,breto2009time} as it bypasses
the explicit calculation of the likelihood function by forward
simulating from the data-generating model.
Unfortunately the application of likelihood--free MCMC is not feasible
in our ``large data'' context. Poor mixing is a
well known problem in inference for diffusion models via MCMC as the
underlying process is by nature very erratic. For large data sets such
as the one exemplified in Section \ref{sec:real-application}
it is very unlikely for a generated trajectory to be close enough
to data to have the corresponding parameter proposal accepted. Low
acceptance rates could be observed even when the value of ${\bm \eta}$
is in the bulk of the posterior distribution support. Another reason
to expect inefficiency in forward simulations is when trajectories are
generated unconditional on data. SMC methods offers an improvement in 
this regards by having the ability to assign larger weights to
particles close to observed data, but we cannot use such approach as explained in the above.

An additional difficulty is related to the simulation of a sufficiently accurate
trajectory for $\{\tau(X_t)\}$. This is in general a
non-issue for SDE models as many approximation schemes are available
\citep{kloeden-platen(1992),rossler(2010)}. In our specific case
numerical discretization is not even required as the process $\{X_t\}$ can
be simulated exactly using the transition densities
(\ref{eq:ou-trans}). Unfortunately, computing $\tau(X_t)$ is not
straightforward because we need to apply the quantile function of the
Gaussian mixture distribution which does not have a closed--form
expression. In practice, we get to solve a nonlinear optimization
problem amounting to finding the zero point $Y_t=\tau(X_t)$ of
$f(\cdot,X_t)=F_{\psi}(\cdot)-\Phi(X_t)$ where $F_{\psi}$ is the
cumulative distribution function defined by
(\ref{eq:mixture-model}). The optimization must be repeated for any
given sampling time $t_i$ ($i=1,...,n$) where in our case $n$ is large
($\approx 25,000$) and for any parameter value ${\bm \eta}$ occuring during
the inferential procedure of choice. This is computationally very
demanding even though the generation of the $\tau(X_t)$'s can be
considered virtually exact, as we control the precision of the
approximated values from the numerical optimization.

Because of the many difficulties highlighted above we revert to
approximate Bayesian computation, which offers a likelihood--free
approach to treat complex stochastic models.

\section{Approximate Bayesian computation}\label{sec:ABC}
The attempt to model complete data sets has dominated the Bayesian
methodology for decades. However, with the advent of large datasets
and complex models this often turns challenging, if not
impossible. Some recent attempts at speeding-up inference via MCMC
using subsets of available data are presented by
\cite{girolami2013playing,korattikara2013austerity} and the
references therein. Aside from the Bayesian framework ``composite
likelihood'' offers several possibilities to simplify computations
with large datasets, see the review in \cite{varin2011overview}.

Approximate Bayesian computation (ABC) offers a principled way to
incorporate information from summary statistics to make
inference for stochastic models for which the likelihood
function is analytically unavailable or computationally too expensive
to approximate, see \cite{marin-et-al(2011)} for a historical
review. Essentially this is done by sampling from an approximation to
the posterior distribution rather than from the exact posterior
distribution itself. 
In the context of our case study, we will show how ABC maintains
essential information about data in a Bayesian procedure while
easing the computational burden considerably. 

Algorithm \ref{alg:pritchard} below summarizes the first genuine ABC
procedure due to \cite{pritchard1999population}. Hereby we
introduce basic notation which is used in the exposition of our own
contribution in section \ref{sec:early-rej-abc}. 
Let $\pi({\bm \eta})$ denote the prior density for ${\bm \eta}$, $p({\bm z}|{\bm \eta})$ the joint
density of the data given ${\bm \eta}$ (i.e. the likelihood function), and
${\bm S}(\cdot)$ a suitable vector of summary statistics, enabling
comparison between a simulated dataset ${\bm z}_{sim}$ and the observed
data ${\bm z}$ according to some measure $\rho(\cdot)$, e.g. the
Euclidean distance,  and the tolerance $\delta\geq 0$.
\begin{algorithm}
\scriptsize
\begin{algorithmic}
\For{$r=1$ to $R$} 
\Repeat 
\State Generate ${\bm \eta}'$ from its prior distribution $\pi({\bm \eta})$
\State Generate ${\bm z}_{sim}$ from the likelihood $p({\bm z}|{\bm \eta}')$
\Until{$\rho({\bm S}({\bm z}_{sim}),{\bm S}({\bm z}))\leq \delta}$
\State set ${\bm \eta}_r={\bm \eta}'$
\EndFor
\end{algorithmic}
\caption{An ABC-rejection algorithm}\label{alg:pritchard}
\normalsize
\end{algorithm}
Algorithm \ref{alg:pritchard} produces $R$ draws from the joint posterior distribution
$\pi({\bm z}_{sim},{\bm \eta}|\rho({\bm S}({\bm z}_{sim}),{\bm S}({\bm z}))\leq \delta)$. When the
generated ${\bm z}_{sim}$ are discarded from the output, the remaining draws are from the ABC marginal posterior of ${\bm \eta}$. Note that when $\delta=0$ and
${\bm S}(\cdot)$ is a sufficient statistic for ${\bm \eta}$, algorithm
\ref{alg:pritchard} samples from the exact posterior
$\pi({\bm z}_{sim},\bm {\eta}|{\bm z})$. On the other hand, when $\delta\rightarrow
\infty$ the algorithm samples from the prior $\pi({\bm \eta})$. In
real life applications, ${\bm S}(\cdot)$ is usually not sufficient and the
choice of a strictly positive $\delta$ must be made in order to make
the procedure computationally feasible. The motivation for ABC is that
an informative summary statistic ${\bm S}(\cdot)$ coupled with a small
tolerance $\delta$ should produce a good approximation to the
exact posterior distribution. Another merit of ABC is that the
likelihood function need not be explicitly known, all that is needed
to run the algorithm is the ability to sample from the
data-generating model.
It is important to notice that ABC methods require careful tuning as both
${\bm S}(\cdot)$, $\rho(\cdot)$ and $\delta$ are user-defined. In
particular, the choice of ${\bm S}(\cdot)$ is
delicate and \cite{fearnhead-prangle(2011)} give directions for constructing
${\bm S}(\cdot)$. A typical choice for $\rho(\cdot)$ is
the uniform kernel, however other possibilities are e.g. the Gaussian and Epanechnikov kernels, see \cite{beaumont2010approximate}.  
We describe the choice of ${\bm S}(\cdot)$ for our case study in Section
\ref{sec:early-rej-abc} below.

\subsection{Early-rejection ABC-MCMC}\label{sec:early-rej-abc}

\begin{figure}
\centering
\includegraphics[height=5cm,width=14cm]{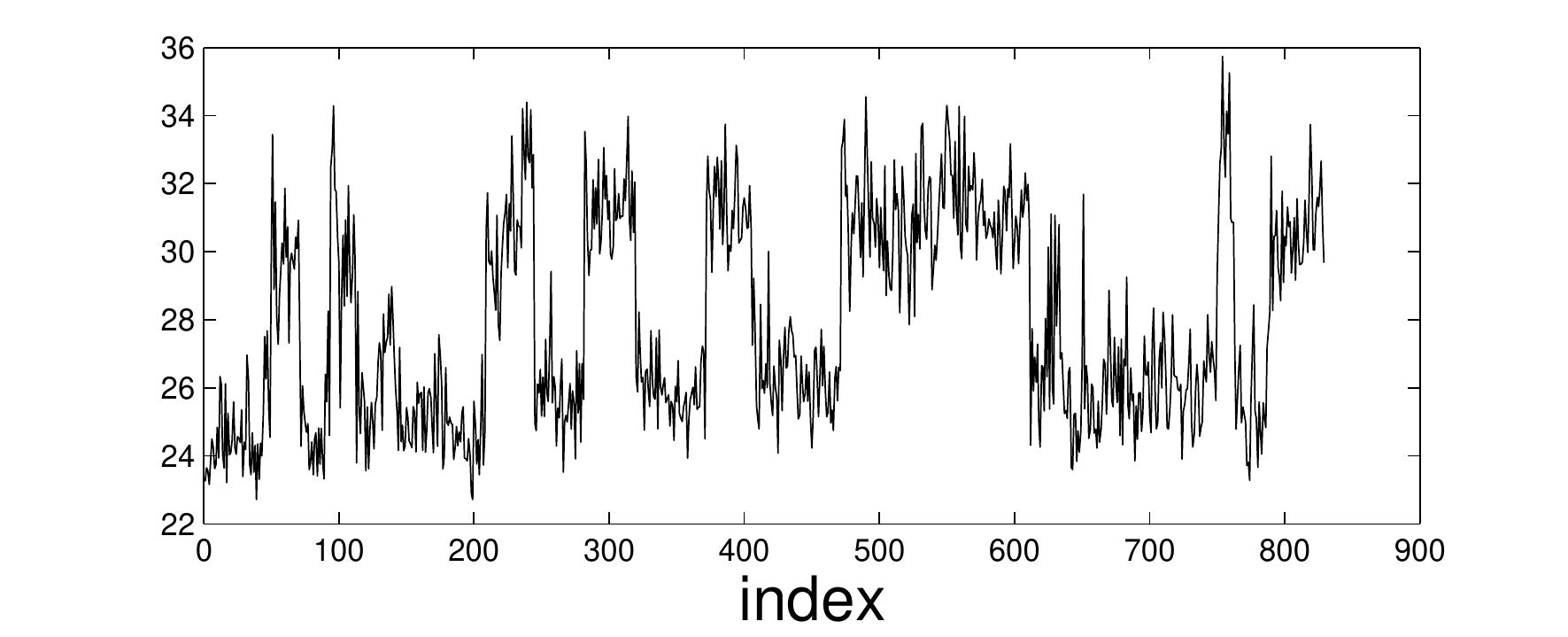} 
\includegraphics[height=5cm,width=9cm]{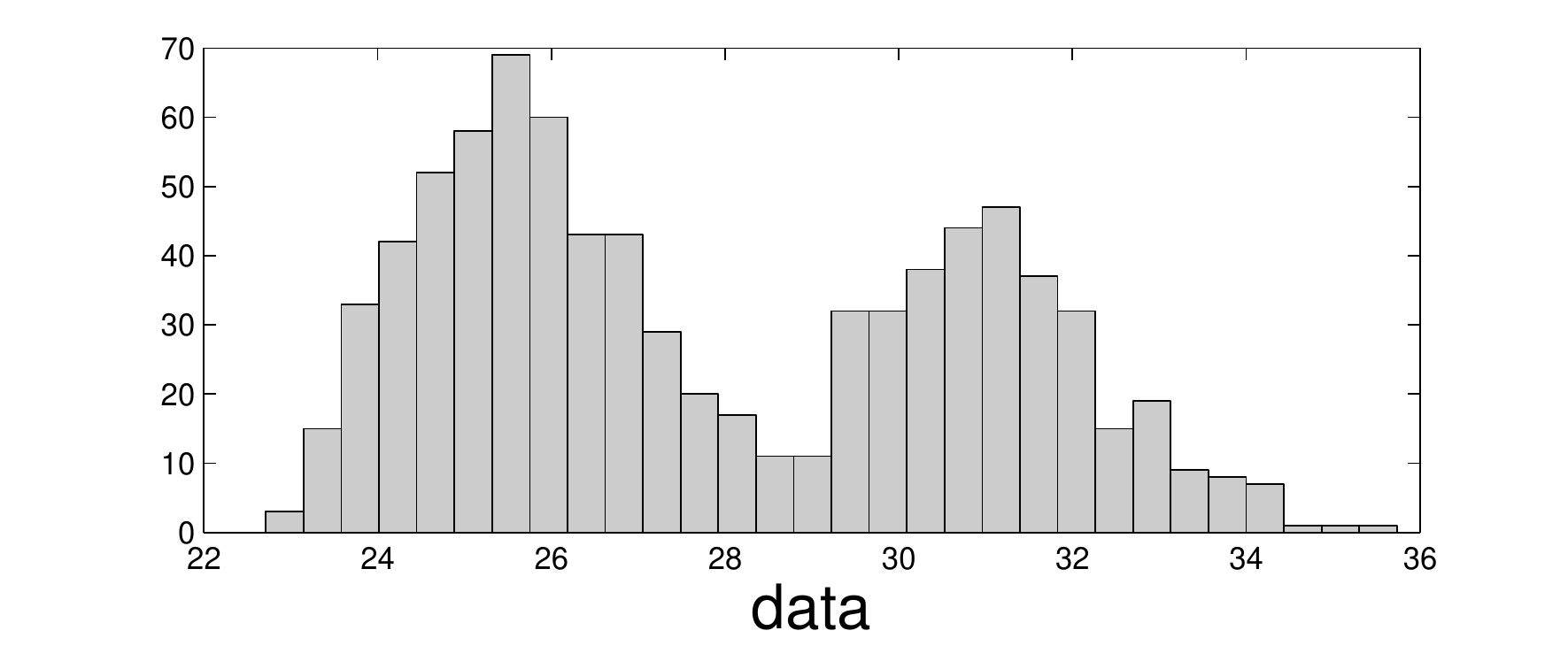} 
\caption{Every 30th observation from the data in Figure \ref{fig:data} is considered here.}
\label{fig:data-reduced}
\end{figure}

Having introduced the basic concepts of ABC, we now
turn to  the ``early-rejection`'' ABC-MCMC algorithm proposed in
\cite{picchini(2012)} and implemented in the \texttt{abc-sde} package
for \textsc{Matlab} \citep{abc-sde}, but with three fundamental
differences. (i) In \cite{picchini(2012)} the vector of summary
statistics ${\bm S}(\cdot)$ was obtained from ``semi-automatic'' regression
following \cite{fearnhead-prangle(2011)}. In particular, the size of
${\bm S}(\cdot)$ was the same as the size of ${\bm \eta}$. In the present case we
use ad-hoc statistics, where $d_s:=\dim({\bm S})$ does not necessarily match
$\dim({\bm \eta})$. (ii) Most importantly, in our application a ``subsample'' of the sampling times
$\{t_{i_1},t_{i_2},...,t_{i_{n'}}\}\subset \{t_0,...,t_n\}$ (with $n'\ll n$) 
is used to simulate trajectories for $\{Z_t\}$. When the times of
subsampling are chosen in a sensible way the features of the model
reflected in the summary statistics are retained while the overall
computational effort is dramatically reduced. As an example of a
subsampling strategy, consider Figure \ref{fig:data-reduced} displaying every $q=30$'th
observation, i.e.\ the $n'=\lceil n/q \rceil$ data at times
$\{t_0,t_{30},t_{60}...,t_{n-30},t_n\}$. Comparing with Figure \ref{fig:data} in which the
complete dataset is displayed, it appears that the qualitative
features of data are preserved by the subsample. Therefore we will
simulate trajectories on a smaller set of times, for example
$\{t_0,t_{30},t_{60},...,t_{n-30},t_n\}$
(in section \ref{sec:real-application} we also experiment with larger values for $n'$). Such procedure leads to summary statistics defined on different sample spaces for real and simulated data, see below. (iii) A user-defined upper bound for $\delta$ is progressively and automatically decreased in our algorithm.

An important question arising in connection with subsampling is how to
choose a set of summary statistics for observed and simulated data.
The latter are produced on a smaller set of time-points and therefore
the comparison between ${\bm S}({\bm z}):\mathbb{R}^n\rightarrow \mathbb{R}^{d_s}$
and ${\bm S}({\bm z}_{sim}):\mathbb{R}^{n'}\rightarrow \mathbb{R}^{d_s}$ is not
immediate. To avoid ambiguity we label the summary
functions corresponding to ${\bm z}$ and ${\bm z}_{sim}$ with ${\bm S}_n$ and ${\bm S}_{n'}$
respectively. Both summary functions must enclose relevant
information for the dynamics of the process as manifested by the
covariance parameters $(\theta,\kappa,\gamma)$
as well as for the static features linked to the parameters of the
stationary distribution ${\bm \psi}=(\mu_1,\mu_2,\sigma_1,\sigma_2,\alpha)$.
For the application
described in section \ref{sec:real-application} we consider
different values of the autocorrelation function of $\{Z_t\}$ 
to represent information pertaining the dynamics of the observed
process. Specifically, we have chosen autocorrelations of the observed
data ${\bm z}$ at lags $(60,300,600,1200,1800,2100)$ and
autocorrelations of ${\bm z}_{sim}$ at lags $(2,10,20,40,60,70)$ when considering $q=30$, so that lags for the subsample match lags for the data (e.g. $2 =60/q$, $10 =300/ q$ etc.). Regarding the marginal
distribution, the summary statistics need not depend on the ordering of the data. We
suggest using empirical percentiles and for our application we choose
the 15th, 30th, 45th, 60th, 75th and 90th empirical percentiles for
the simulated data ${\bm z}_{sim}$ to be compared with the corresponding
percentiles for the observed data ${\bm z}$. 

Algorithm \ref{alg:lfmcmc-earlyrej} reports our ABC-MCMC
procedure. The algorithm proposes simultaneously draws for ${\bm \eta}$ and
$\delta$ with the purpose of retrospectively filtering-out 
%(once $R$ draws of $({\bm \eta},\delta)$ have been obtained) 
the ${\bm \eta}$'s by retaining only those corresponding to sufficiently
small $\delta$'s. Further, the algorithm is often able to
``early-reject'' proposed draws without having to generate ${\bm z}_{sim}$
due to our choice of a uniform kernel for $\rho(\cdot)$; i.e. we set\
$$\rho({\bm S}_{n'}({\bm z}_{sim}),{\bm S}_n({\bm z}))=K \left(\frac{|{\bm S}_{n'}({\bm z}_{sim})-{\bm S}_n({\bm z})|}{\delta}\right)$$
where  
\begin{equation}
\label{eq:unikernel}
K({\bm w})= I({\bm w}\colon\,{\bm w}^T \mathbf{A}{\bm w} < \pi^{-1}(\Gamma(d_s/2)d_s/2)^{2/d_s}|\mathbf{A}|^{1/d_s}).
\end{equation}
The $\pi$ in \eqref{eq:unikernel} denotes the mathematical constant, $I(\cdot)$ is the indicator function and $\mathbf{A}$
is a user-defined $d_s \times d_s$ diagonal matrix of positive weights scaling the values of the entries in the vector of summary statistics. Notice that the quantity on the right hand side of the inequality in \eqref{eq:unikernel} is the unique value such that the volume of the region ${\bm w}^T \mathbf{A}{\bm w}$ equals 1. We refer to
\cite{fearnhead-prangle(2011),picchini(2012)} for additional details.
%$c=V|A|^{1/d_s}$, where $V=%\pi^{-1}(\Gamma(d_s/2)d_s/2)^{2/d_s}$ and 
% In our case $w=|{\bm S}_{n'}({\bm z}_{sim})-{\bm S}_n({\bm z})|/\delta$. 
We can initially check whether to reject
the proposed  $({\bm \eta}',\delta')$ by evaluating a part of the
traditional Metropolis-Hastings acceptance ratio; the one denoted as
``ratio'' in algorithm \ref{alg:lfmcmc-earlyrej} below. When the
draw $\omega\sim U(0,1)$ from the uniform distribution in step 2 is
larger than this ratio we can immediately reject the proposed
parameters regardless the value of $K(\cdot)\in \{0,1\}$ (which in fact does not need to be computed at this stage) and
\textit{without} having to simulate ${\bm z}_{sim}$. When $\omega$ is smaller or equal than ``ratio'' ${\bm z}_{sim}$ is produced and the usual Metropolis-Hastings procedure is resumed. This is extremely
beneficial from the computational point of view, especially since ABC
methods are usually performed at low acceptance rates. Another ``early
rejection'' mechanism for ABC has been suggested in the ``one-hit MCMC-ABC''
algorithm by \cite{lee-andrieu}. Notice in step 1 of the algorithm the proposal mechanism for ${{\bm \eta}}$ and $\delta$ is written in a very general way: however in our experiments we assume the two quantities to be independent and therefore we could
also write $u({{\bm \eta}},\delta|{{\bm \eta}}_r,\delta_r) = u_1 ({{\bm \eta}}|{{\bm \eta}}_r)u_2(\delta|\delta_r)$ with $u_1(\cdot)$ and  $u_2(\cdot)$ the corresponding proposal distributions.
For $u_1(\cdot)$ we employ an automatically tuned
Metropolis random walk with Gaussian innovations
\cite{haario-et-al(2001)}. Therefore in practice $u_1$ is used to
simulate log-transformed parameters
${\bm \eta}=(\log\theta,\log\kappa,\log\gamma,\log\mu_1,\log\mu_2,\log\sigma_1,\log\sigma_2,\log\alpha)$. For $u_2$ we consider a
(truncated) Gaussian Metropolis random walk on the support
$(-\infty,\log\delta_{max}]$ where $\delta_{max}$ is initially set by the user and during the algorithm execution it gets automatically decreased until a user defined threshold $\delta_{minmax}$ is reached (see the \texttt{update} step in algorithm \ref{alg:lfmcmc-earlyrej}). In our experiments the update procedure for $\delta_{max}$ is executed every $g=3,000$ ABC-MCMC iterations using $m=99$, i.e.  $\delta_{max}$ is assigned the 99th percentile from the last $g$ simulated values $\delta_{(l-1)g:lg-1}:=(\delta_{(l-1)g},\delta_{(l-1)g+1},...,\delta_{lg-1})$, for $l=1,2,...$, see also \cite{lenormand2013adaptive}. If the percentile is smaller than $\delta_{minmax}$ we set $\delta_{max}:=\delta_{minmax}$. This way the algorithm does not waste computational time for simulations corresponding to excessively large values of $\delta$. Of course the choice of $\delta_{max}$ and $\delta_{minmax}$ is applications specific and has to be a balanced compromise between exploration of the posterior surface (not too small $\delta_{max}$ nor $\delta_{minmax}$) and inferential accuracy (not too large $\delta_{max}$ and $\delta_{minmax}$).   

\begin{algorithm}
\footnotesize
\begin{algorithmic}
\State 0. Initialization: Compute ${\bm S}_n({\bm z})$. Fix $R$, $m$, $g$, $\delta_{minmax}<\delta_{max}$ and $\delta_{start}\leq\delta_{max}$. 
Simulate ${{\bm \eta}}_{start}\sim \pi({{\bm \eta}})$, 
${{\bm x}}_{start}\sim\pi({{\bm x}}|{{\bm \eta}}_{start})$, and
${\bm z}_{start}\sim \pi({\bm z}|\tau({\bm x}_{start}),{\bm \eta}_{start})$. 
Set $r=0$, $({{\bm \eta}}_0,\delta_0)\equiv({{\bm \eta}}_{start},\delta_{start})$, and 
${{\bm S}}_{n'}({{\bm z}}_{sim,0})\equiv {{\bm S}}_{n'}({{\bm z}}_{start})$. \\ \\
%such that $K(|{{\bm S}}_{n'}({{\bm z}}_{start})-{{\bm S}}_n({{\bm z}})|/\delta_{start})\equiv 1$.

At $(r+1)$th ABC-MCMC iteration:\\
\State 1.  generate $({\bm \eta}',\delta')\sim
u({\bm \eta},\delta|{{\bm \eta}}_r,\delta_r)$ and \texttt{update} $\delta_{max}$
if appropriate (see $\star$ below);
\State 2. generate $\omega \sim U(0,1)$;
\If{\[\omega>\frac{\pi({{\bm \eta}}')\pi(\delta')u({{\bm \eta}}_r,\delta_r|{{\bm \eta}}',\delta')}{\pi({{\bm \eta}}_r)\pi(\delta_r)u({{\bm \eta}}',\delta'|{{\bm \eta}}_r,\delta_r)}\qquad (=\text{``ratio''}) \]}
\State $({{\bm \eta}}_{r+1},\delta_{r+1},{{\bm S}}_{n'}({{\bm z}}_{sim,r+1})):=({{\bm \eta}}_r,\delta_r,{{\bm S}}_{n'}({{\bm z}}_{sim,r}))$; \Comment{(proposal early-rejected)}
\Else{ generate ${{\bm x}}_{sim}\sim \pi({{\bm x}}|{{\bm \eta}}')$ and ${{\bm z}}_{sim}\sim\pi({{\bm z}}|\tau({\bm x}_{sim}),{{\bm \eta}}')$ conditionally on the  ${{\bm \eta}}'$ from step 1}
%and calculate ${{\bm S}}_{n'}({{\bm z}}_{sim})$; 
   \If{$K(|{{\bm S}}_{n'}({{\bm z}}_{sim})-{{\bm S}}_n({{\bm z}})|/\delta')=0$} 
      \State $({{\bm \eta}}_{r+1},\delta_{r+1},{{\bm S}}_{n'}({{\bm z}}_{sim,r+1})):=({{\bm \eta}}_r,                
      \delta_r,{{\bm S}}_{n'}({{\bm z}}_{sim,r}))$ \Comment{(proposal rejected)}
    \ElsIf{$\omega\leq \mathrm{ratio}$} 
      \State $({{\bm \eta}}_{r+1},\delta_{r+1},{{\bm S}}_{n'}({{\bm z}}_{sim,r+1})):=({{\bm \eta}}',   
      \delta',{{\bm S}}_{n'}({{\bm z}}_{sim}))$ \Comment{(proposal accepted)}
     \Else
     \State $({{\bm \eta}}_{r+1},\delta_{r+1},{{\bm S}}_{n'}({{\bm z}}_{sim,r+1})):=({{\bm \eta}}_r,        
      \delta_r,{{\bm S}}_{n'}({{\bm z}}_{sim,r}))$ \Comment{(proposal rejected)}
      \EndIf

\EndIf
\State 3. increment $r$ to $r+1$. If $r>R$ stop, else go to step 1.
\\
\\
$^\star$ This is the \texttt{update} procedure for $\delta_{max}$:\\
When iteration $r$ is a multiple of a positive integer $g$, i.e. $r=l\cdot g$ for $l=1,2,...$
set $\delta_{max}$ as the $m$th percentile of $\delta_{(l-1)g:r-1}$. If $\delta_{max}<\delta_{minmax}$ set $\delta_{max}:=\delta_{minmax}$.
\end{algorithmic}
\caption{Early--Rejection ABC-MCMC}
\label{alg:lfmcmc-earlyrej}
\normalsize
\end{algorithm}

Note that whenever we write ${\bm x}_{sim}\sim \pi({\bm x}|{\bm \eta'})$ it means that we are simulating the Markov process $\{X_t\}$ conditionally on some ${\bm \eta'}$ using \eqref{eq:ou-trans} and starting at $X_0=x_0$, where $x_0$ is a constant determined in section \ref{sec:real-application}. Once ${\bm x}_{sim}$ is available we apply the $\tau(\cdot)$ transformation to obtain $\tau({\bm x}_{sim})$ and then add a realization of $\{U_t\}$ (generated using its own transition density and of course conditionally on ${\bm \eta'}$). The result is a realization of what is synthetically denoted with ${{\bm z}}_{sim}\sim\pi({{\bm z}}|\tau({\bm x}_{sim}),{{\bm \eta}}')$, that is a realization of process $\{Z_t\}$.

All trajectories are generated at times belonging to the subsample $\{t_{i_1},t_{i_2},...,t_{i_{n'}}\}$, i.e. ${\bm z}_{sim}=(z_{sim,t_{i_1}},...,z_{sim,t_{i_{n'}}})$ (and similarly for ${\bm x}_{sim}$) and the corresponding ${\bm S}_{n'}({\bm z}_{sim})$ is then compared to the statistics for the full dataset ${\bm S}_n({\bm z})$. 
Also notice that conditional independence of observations is nowhere invoked in algorithm  \ref{alg:lfmcmc-earlyrej}, which is therefore suitable for the diffusion model with error \eqref{eq:state-space}.
Algorithm \ref{alg:lfmcmc-earlyrej} produces $R$ draws $\{{\bm \eta}_r,\delta_r\}_{r=1:R}$ from the augmented posterior $\pi({\bm \eta},\delta|\rho({\bm S}({\bm z}_{sim}),{\bm S}({\bm z}))\leq \delta)$ but we are only interested in the marginal posterior $\pi({\bm \eta}|\rho({\bm S}({\bm z}_{sim}),{\bm S}({\bm z}))\leq \delta)$: therefore once the algorithm run has been completed we filter-out draws for ${\bm \eta}$ which are not consistent with some suitable (small enough) threshold $\delta^*$. A strategy for ``filtering'' the output and determining $\delta^*$ is illustrated in section \ref{sec:simulation-study}, see also \cite{picchini(2012),bortot2007inference}.

\section{Simulation study: a comparison with exact Bayesian inference}
\label{sec:simulation-study}
We have conducted a small-sample simulation study to compare
results from our ABC-MCMC algorithm with exact Bayesian inference
based on the particle MCMC methodology \cite{andrieu2010particle}
in form of a parallelised version proposed in \cite{drovandi(2013)}. 

Particle MCMC produces exact Bayesian inference whenever an unbiased
estimate $\hat{p}({\bm z}|{\bm \eta})$ to the likelihood in
\eqref{eq:full-likelihood} can be computed. This is possible for model
(\ref{eq:state-space}) as explained in what follows. Note that
conditionally on the latent state
$\{\tau(X_{0})=\tau(x_{0}),\ldots,\tau(X_{{j}})=\tau(x_{{j}})\}$, the
observation $(z_{0},\ldots,z_{{j}})$ is merely a translation of the
measurement errors thus having density
$p(z_{0},\ldots,z_{{j}}|\tau(x_{0}),\ldots,\tau(x_{{j}});{\bm \eta})$ equal to 
$$
\frac{1}{\gamma}\cdot\phi\left(\frac{z_{{0}}-\tau(x_{{0}})}{\gamma}\right)
\times\prod_{i=1}^j\frac{1}{\gamma\sqrt{1-e^{-2\kappa\Delta_i}}}\cdot
\phi\left(\frac{z_{i}-\tau(x_{i})-e^{-\kappa\Delta_i}(z_{{i-1}}-\tau(x_{{i-1}}))}
{\gamma\sqrt{1-e^{-2\kappa\Delta_i}}}\right),
$$
where $\Delta_i=t_i-t_{i-1}$ and where $\phi(\cdot)$ denotes the density of the standard Gaussian
distribution. We obtain an approximation to $p(z_0,...,z_j|{\bm \eta})$ via
SMC by use of the bootstrap filter of \cite{gordon1993novel}, see also
\cite{doucet2001sequential}. Let $\{\tau(x_{i-1}^l)\}$ denote the set of $N$
particles available at time $t_{i-1}$ \textit{before randomisation
  occur}, and $\{\tau(\tilde{x}_{i-1}^l)\}$ the resulting randomised
particles which are used as a starting point to propagate particles
forward to time $t_i$. Then
$$\hat{p}(z_0,...,z_{j}|{\bm \eta})=\prod_{i=1}^{j}\hat{p}(z_i|z_{0},...,z_{i-1};{\bm \eta})
=\prod_{i=1}^{j}\frac{1}{N}\sum_{l=1}^Nw_i^l$$
with weights $w_i^l$ ($l=1,...,N$; $i=1,...,n$) given by
\begin{equation}
w_i^l = \frac{1}{\gamma\sqrt{1-e^{-2\kappa\Delta_i}}}\cdot
\phi\left(\frac{z_i-\tau(x_{i}^l)-e^{-\kappa\Delta_i}(z_{i-1}-\tau(\tilde{x}_{i-1}^l))}
{\gamma\sqrt{1-e^{-2\kappa\Delta_i}}}\right).\label{eq:smc-weights}
\end{equation}
Note that $w_i^l$ depend on $\tau(\tilde{x}_{i-1}^l)$ which is the parent of
$\tau(\tilde{x}_{i}^l)$ in the genealogy of the $l$'th particle. 
Finally we can compute the (unbiased) likelihood approximation
\begin{align*}
\hat{p}({\bm z}|{\bm \eta})&=\hat{p}(z_0|{\bm \eta})\prod_{i=1}^n\hat{p}(z_i|z_{0},...,z_{i-1};{\bm \eta})\\
\text{where}\quad \hat{p}(z_0|{\bm \eta}) &= \frac{\sum_{l=1}^Nw_0^l}{N}, \quad w_0^l=\frac{1}{\gamma}\cdot\phi\biggl(\frac{z_0-\tau(\tilde{x}_{0}^l)}{\gamma}\biggr).
\end{align*}
The procedure above can be parallelised over $M$ machines/cores to
obtain $M$ independent approximations of $p({\bm z}|{\bm \eta})$ for the running
value of ${\bm \eta}$. The average of these approximations is a more precise (unbiased) estimate of the
likelihood which can be used in the Metropolis-Hastings procedure to
produce exact Bayesian inference for ${\bm \eta}$.  Parallel computation
improves the mixing of the resulting chain for particle MCMC, although
only marginally for a small $M$. We used the \texttt{parfor}
functionality from the Parallel Computing Toolbox for \textsc{Matlab}
(release R2013a) with $M=4$ cores and $N=100$ particles for each core.

As mentioned in section \ref{sec:issues-exact-bayesian}, running an
exact Bayesian algorithm based on SMC on a large dataset is extremely
time consuming when considering a model such as
\eqref{eq:state-space}. This would be the case with the sample size
$n=24,842$ of the data in our application, section \ref{sec:real-application}.
Therefore we conduct a simulation study with artificial data of a much
smaller size. As model parameters we used the parameters denoted with
``true values'' in Table \ref{tab:estimates-simulated-data}. Setting
the initial state to $x_0=-2.45$ we produced $n=355$ observations from
model \eqref{eq:state-space} at times $\{1,71,141,...,24781\}$. The simulated data (not reported) have switching structure resembling Figure \ref{fig:data-reduced}. 
Please note that in this case we are not making use of subsampling as the $n=n'=355$
data points are considered to be a full dataset. Therefore, ABC
and exact Bayesian results are based on the same amount of data. A
proper subsampling experiment is considered in section \ref{sec:real-application}. 

\begin{figure}
\centering
\includegraphics[scale=0.5]{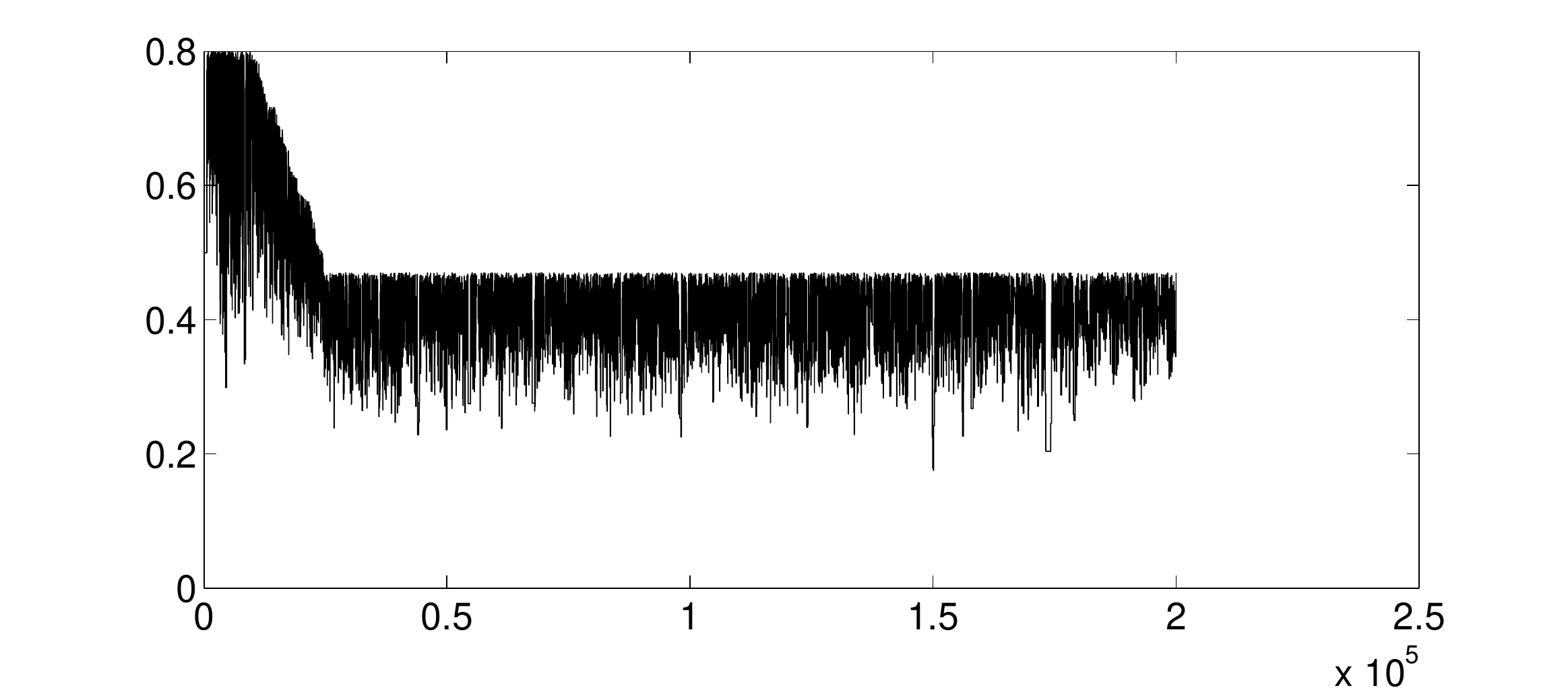}
\caption{Thinned chain for $\delta$ for the simulated data problem.}\label{fig:bandwidth-simuldata}
\end{figure}

We employ the following uniform priors: $\log\theta \sim U(-7,-5.3),$
$\log\kappa\sim U(-1.5,0.3)$,  $\log\gamma\sim U(-0.7,0.5)$, $\log \mu_1\sim U(3.1,3.3)$,
$\log\mu_2\sim U(3.3,3.7)$, $\log\sigma_1 \sim U(-2.5,1)$,
$\log\sigma_2\sim U(-2.5,1)$, $\log\alpha  \sim U(-1.5,-0.05)$. 
The ABC summary statistics comprise autocorrelation values at lags 2,
5, 10 and 15 together with the 15, 30, 45, 60, 75, 90th percentiles
for both ${\bm z}$ and ${\bm z}_{sim}$. Hence, both ${\bm S}_n(\cdot)$ and
${\bm S}_{n'}(\cdot)$ have length $d_s=10$. Algorithm
\ref{alg:lfmcmc-earlyrej} was run for $R=2\times 10^6$ iterations,
with starting bandwidth $\delta_{start}=0.5$ and exponential prior on $\delta\sim
Exp(0.2)$. The proposals for $\log\delta$ were generated via
(truncated) Gaussian Metropolis random walk on the support
$(-\infty,\log\delta_{max}]$ using steps having variance 0.2. We update
the initial $\delta_{max}=0.8$ as described in section \ref{sec:early-rej-abc} and using $\delta_{minmax}=0.47$. The
weight matrix $\mathbf{A}$ defining the uniform kernel (\ref{eq:unikernel})
was set to $\mathrm{diag}(\mathbf{A})=[100,100,100,100,1,1,1,1,1,1]$. This assigns larger
weights to the autocorrelations to compensate for their smaller values
compared to the percentiles. Results were obtained in about 4.7 hrs on
a Intel Core i7-2600 CPU 3.40 GhZ with 4 Gb RAM. We observed an
acceptance rate in the range 0.3--1\% during the simulations, which is
a good compromise between statistical accuracy (the smaller $\delta$
the larger the rejection rate) and exploration of the posterior
surface. We thinned the generated chain by retaining each 10th draw
and then removed as burnin the first 30,000 draws, essentially disregarding
draws corresponding to the update phase for $\delta_{max}$ and $\delta_{minmax}$, see
Figure \ref{fig:bandwidth-simuldata}. 

\begin{figure}
\centering
\includegraphics[scale=0.5]{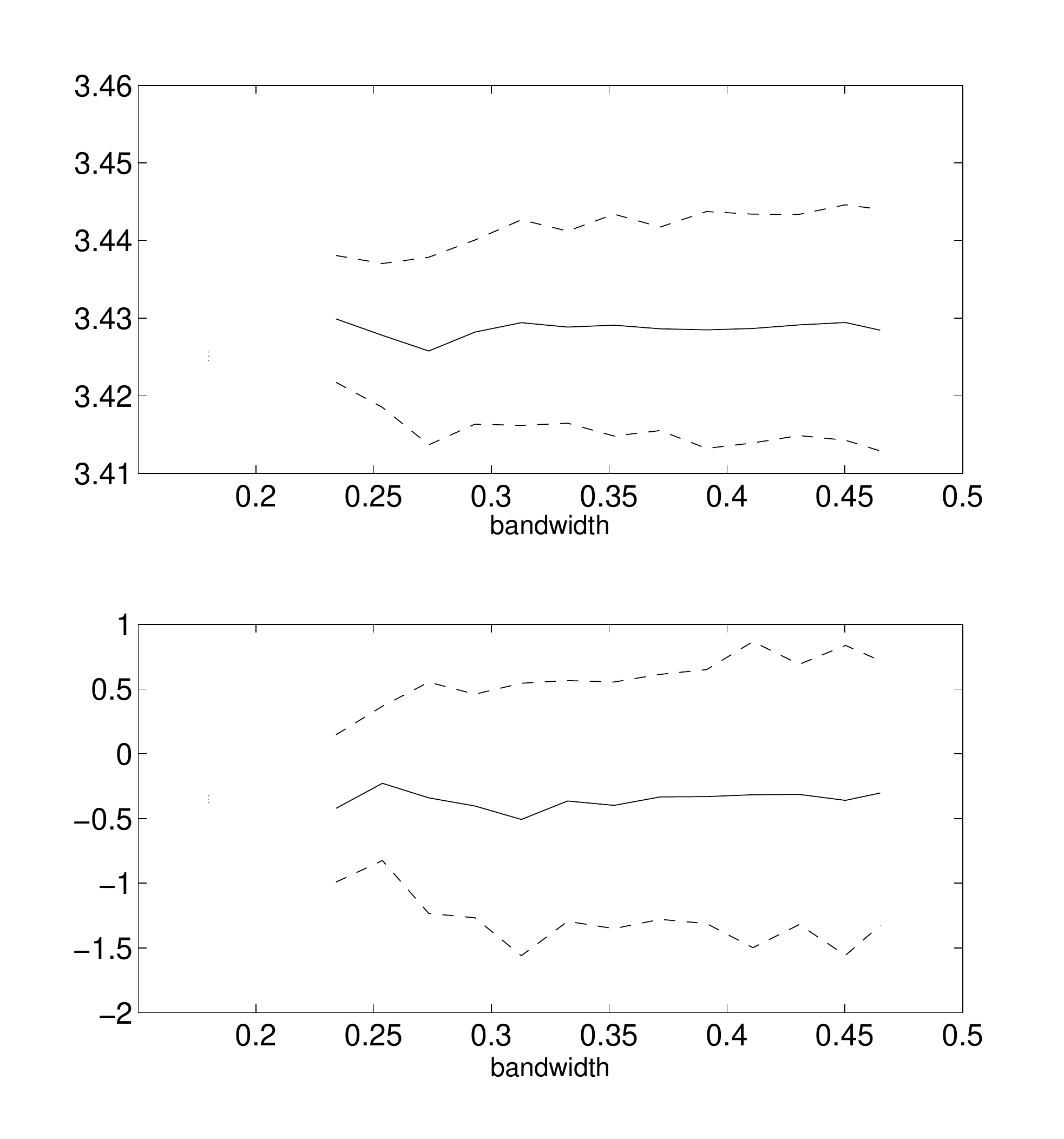}
\caption{Marginal posterior means vs $\delta$ before filtering the ABC-MCMC output [$\pm 2$ SD]  for $\log \mu_2$ (top) and $\log \sigma_2$ (bottom).}\label{fig:posteriormeans_vs_bandwidth}
\end{figure}

Finally by inspecting plots as in Figure \ref{fig:posteriormeans_vs_bandwidth} we ``filtered'' the
remaining chain by studying the posterior means for varying values of
$\delta$ and ultimately selected draws for ${\bm \eta}$ corresponding to
$\delta$'s not exceeding $\delta^*=0.35$, where $\delta^*$ has been defined at the end
of section \ref{sec:early-rej-abc}. Note that this is possible as our
ABC-MCMC algorithm produces chains for both ${\bm \eta}$ and
$\delta$. Inferential results from the remaining 28,000 draws are
compared to particle MCMC (exact Bayesian inference) in Table \ref{tab:estimates-simulated-data}.
The particle MCMC algorithm was run for $R=200,000$
iterations. Results were obtained in about 67 hrs, with an average
acceptance rate of 10\%. After removing the initial 25,000 draws
(burn-in) we produced the exact (up to Monte Carlo sampling)
inferential results given in Table \ref{tab:estimates-simulated-data}. 

\begin{table}
\centering 
\caption{Posterior means and 95\% posterior intervals from particle
  MCMC (lines without asterisks) and ABC-MCMC (lines with asterisks) when $n=355$.}
\begin{tabular}{cccl}
\hline
\hline
{} & True values &  {} & {}\\ 
\hline
$\log\theta$   & $-5.914$ & $-6.108$ & $[-6.634,-5.700]$  \\
{}         & {}     & $-6.244^*$ &$[ -6.744,-5.738]$\\
$\log\kappa$   & $-0.620$  & $-0.811$  &$[-1.473,0.206]$ \\
{}         & {}     & $-0.902^*$ &$[-1.460,0.014]$\\
$\log\gamma$   & 0.061  & 0.072 & $[-0.040,0.171]$\\
{}         & {}     & $-0.002^*$& $[-0.233,0.219]$\\
$\log\mu_1$    & 3.24  & 3.24 & $[3.24,3.25]$\\
{}         & {}     & $3.25^*$ &$[3.23,3.26]$\\
$\log\mu_2$    & 3.43  & 3.43 & $[3.42,3.43]$\\
{}         & {}     & $3.43^*$& $[3.42,3.44]$\\
$\log\sigma_1$ & $-0.616$  & $-0.401$ & $[-0.803,-0.077]$\\ 
{}         & {}     & $-1.586^*$ &$[-2.359,-1.088]$\\
$\log\sigma_2$ & $-0.472$  & $-0.852$ & $[-1.936,-0.190]$\\ 
{}         & {}     & $-0.392^*$ &$[-1.379,0.373]$\\
$\log\alpha$   & $-0.622$  & $-0.652$ & $[-0.970,-0.426]$\\ 
{}         & {}     & $-0.630^*$ &$[-0.916,-0.423]$\\
\hline
\end{tabular}
\label{tab:estimates-simulated-data} 
\end{table}

\begin{figure}
\centering
\subfigure[$\log \theta$]{\includegraphics[scale=.35]{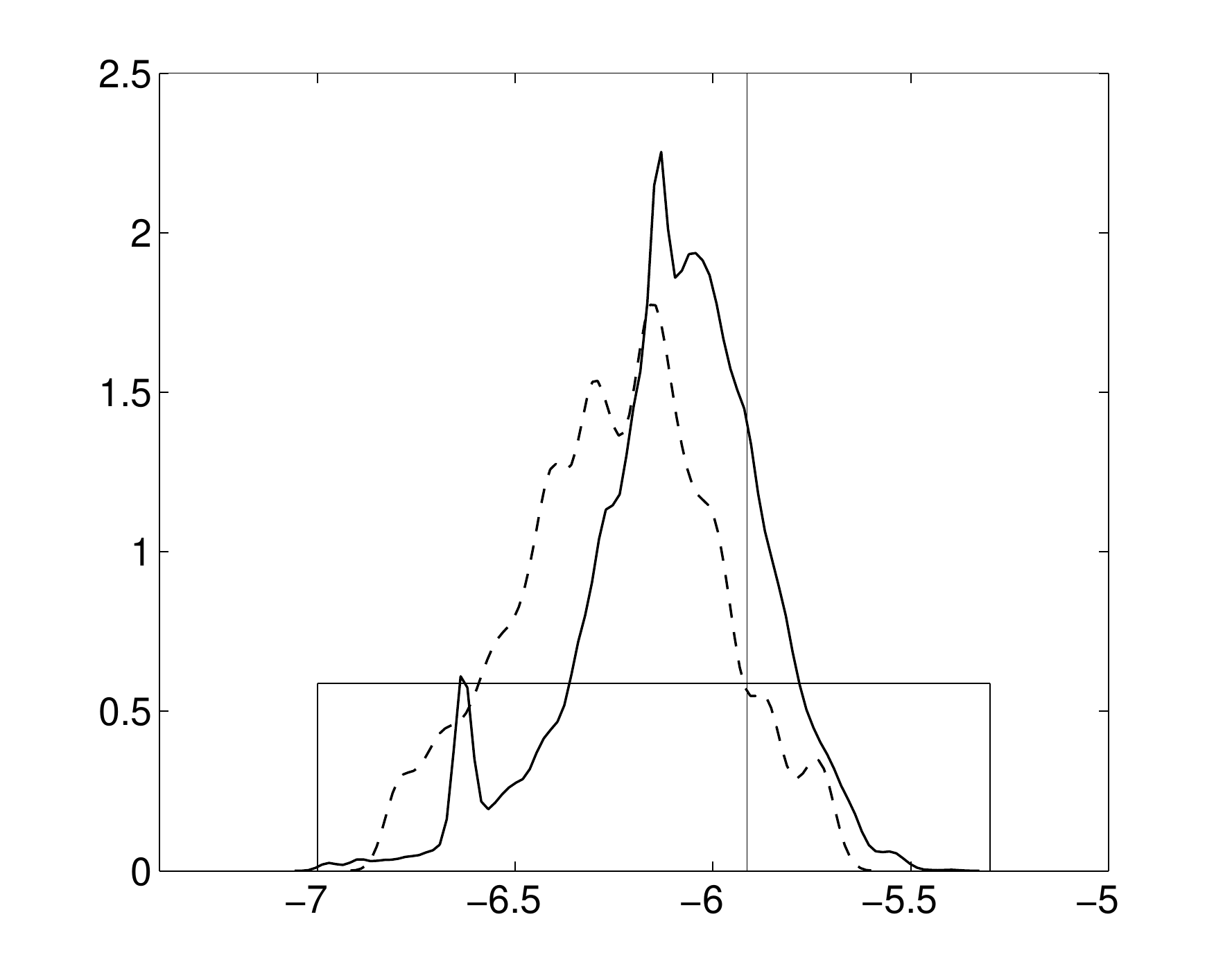}}
\subfigure[$\log \kappa$]{\includegraphics[scale=.35]{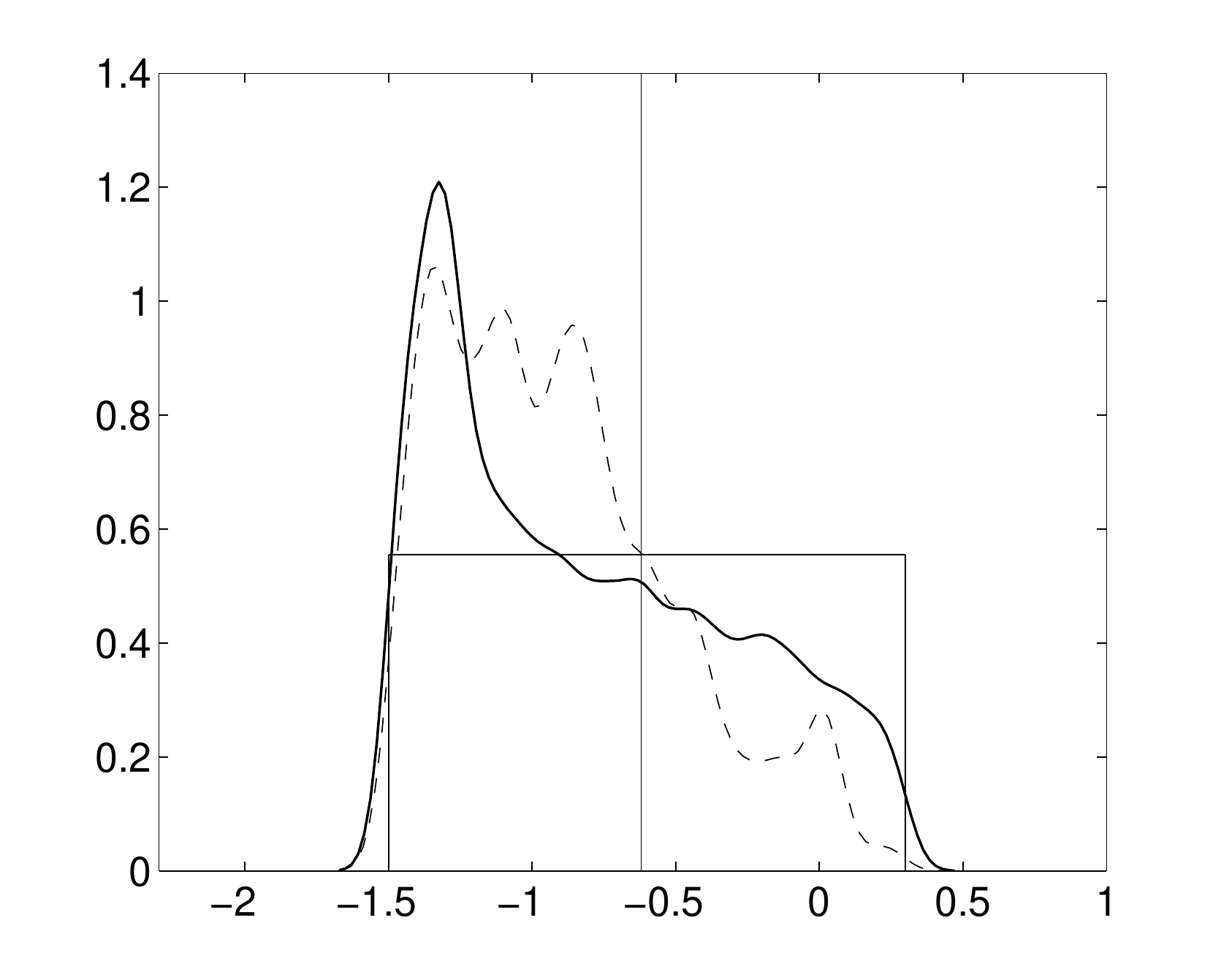}}\\
\subfigure[$\log \gamma$]{\includegraphics[scale=.35]{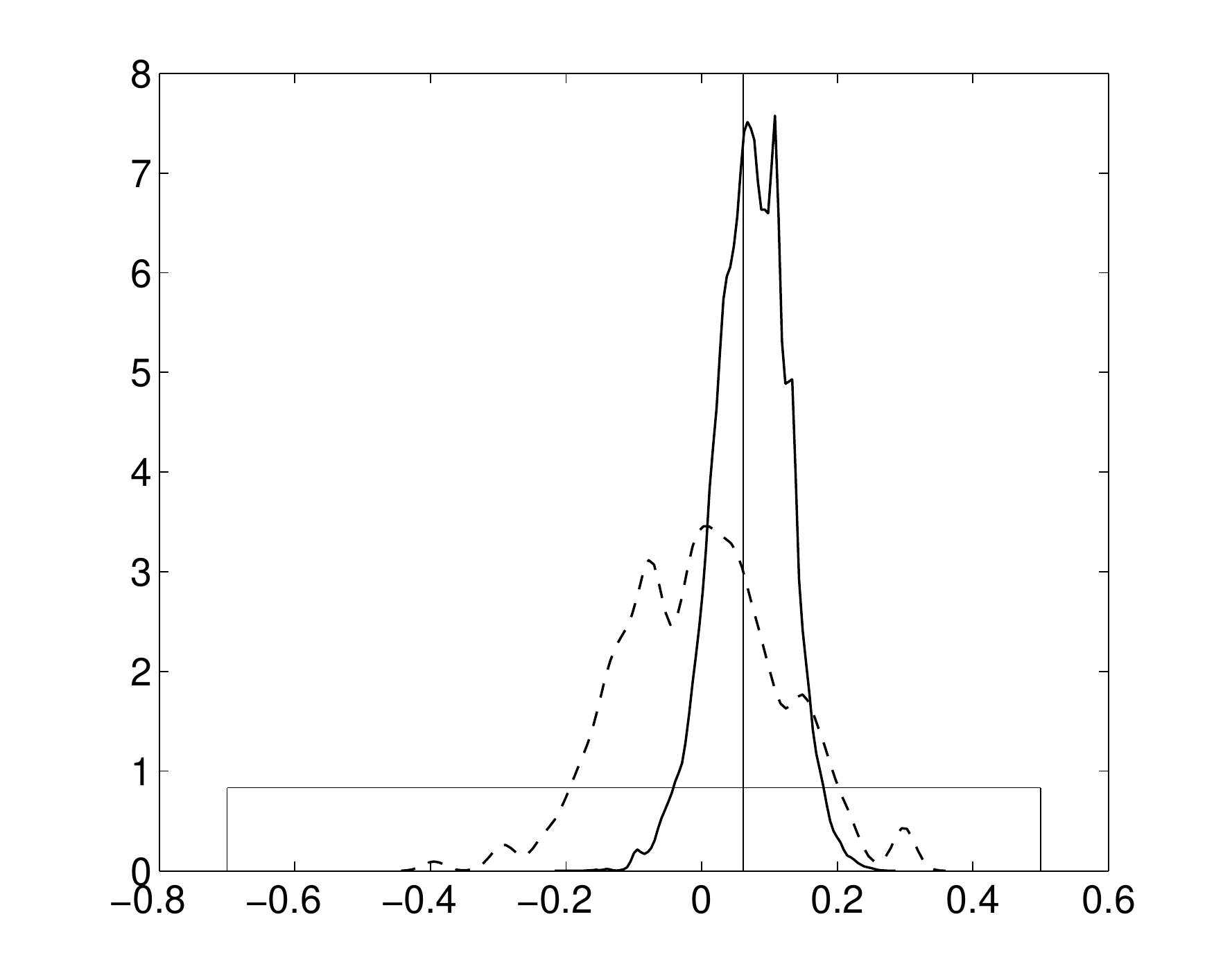}}
\subfigure[$\log \alpha$]{\includegraphics[scale=.35]{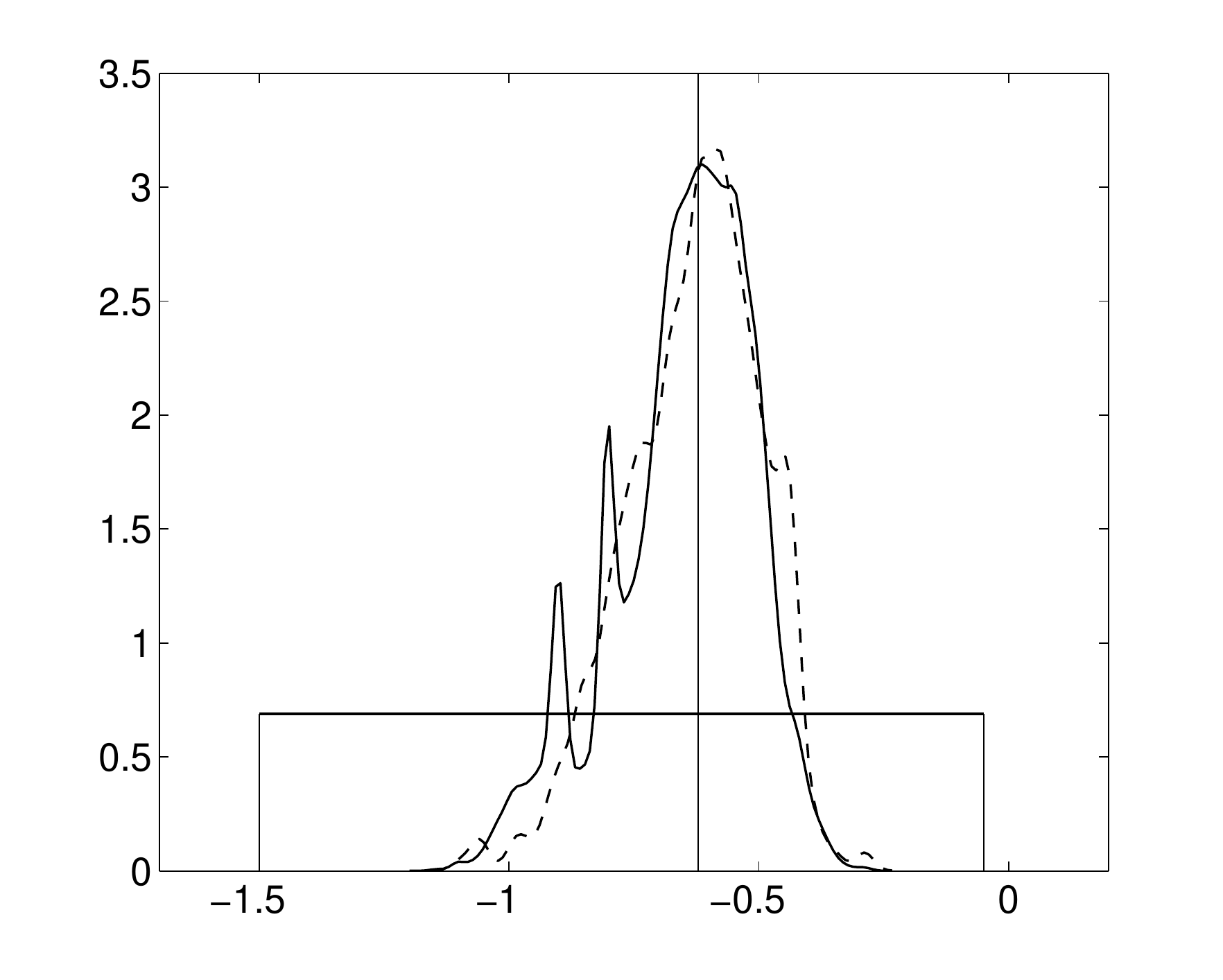}}\\
\subfigure[$\log \mu_1$]{\includegraphics[scale=.35]{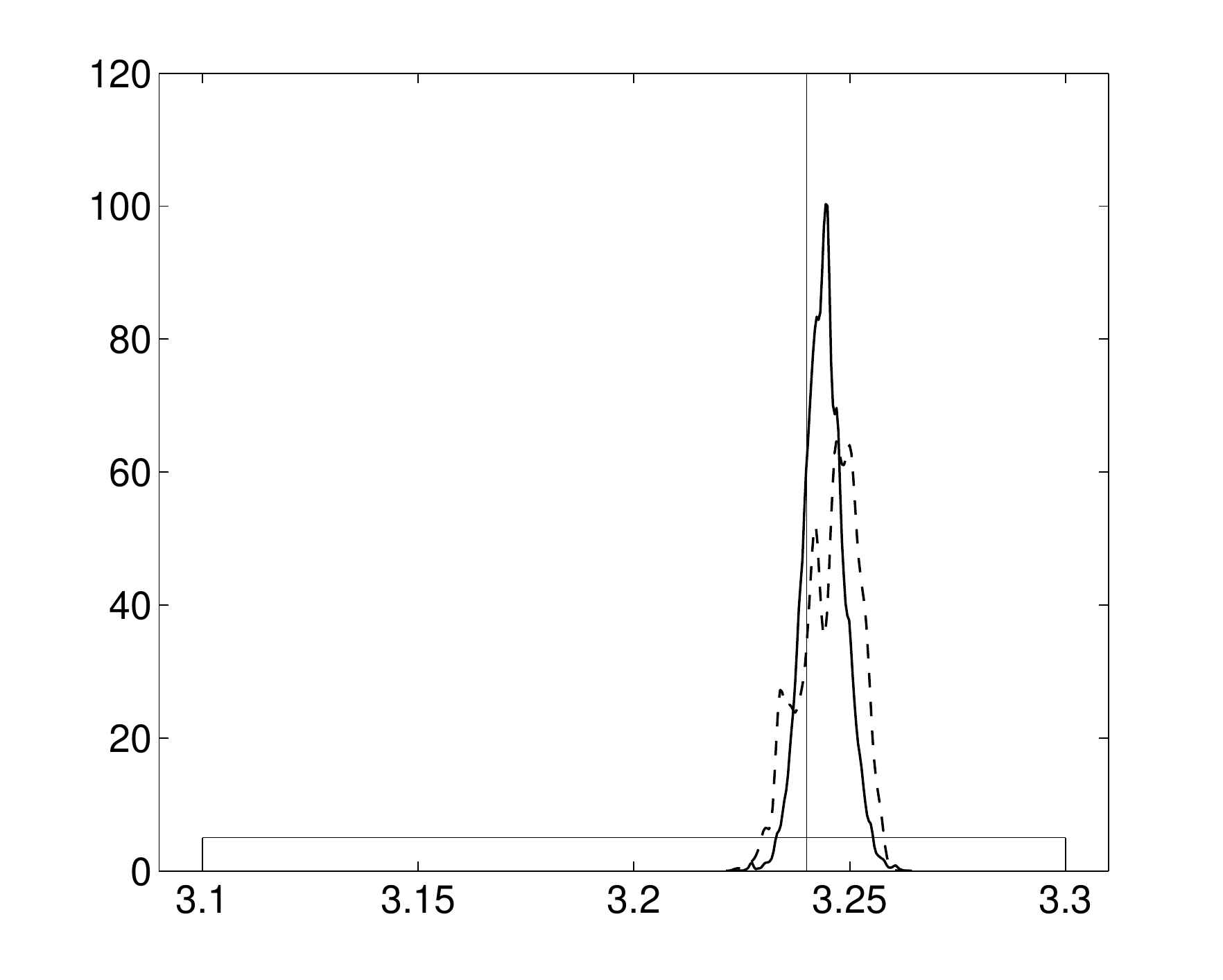}}
\subfigure[$\log \mu_2$]{\includegraphics[scale=.35]{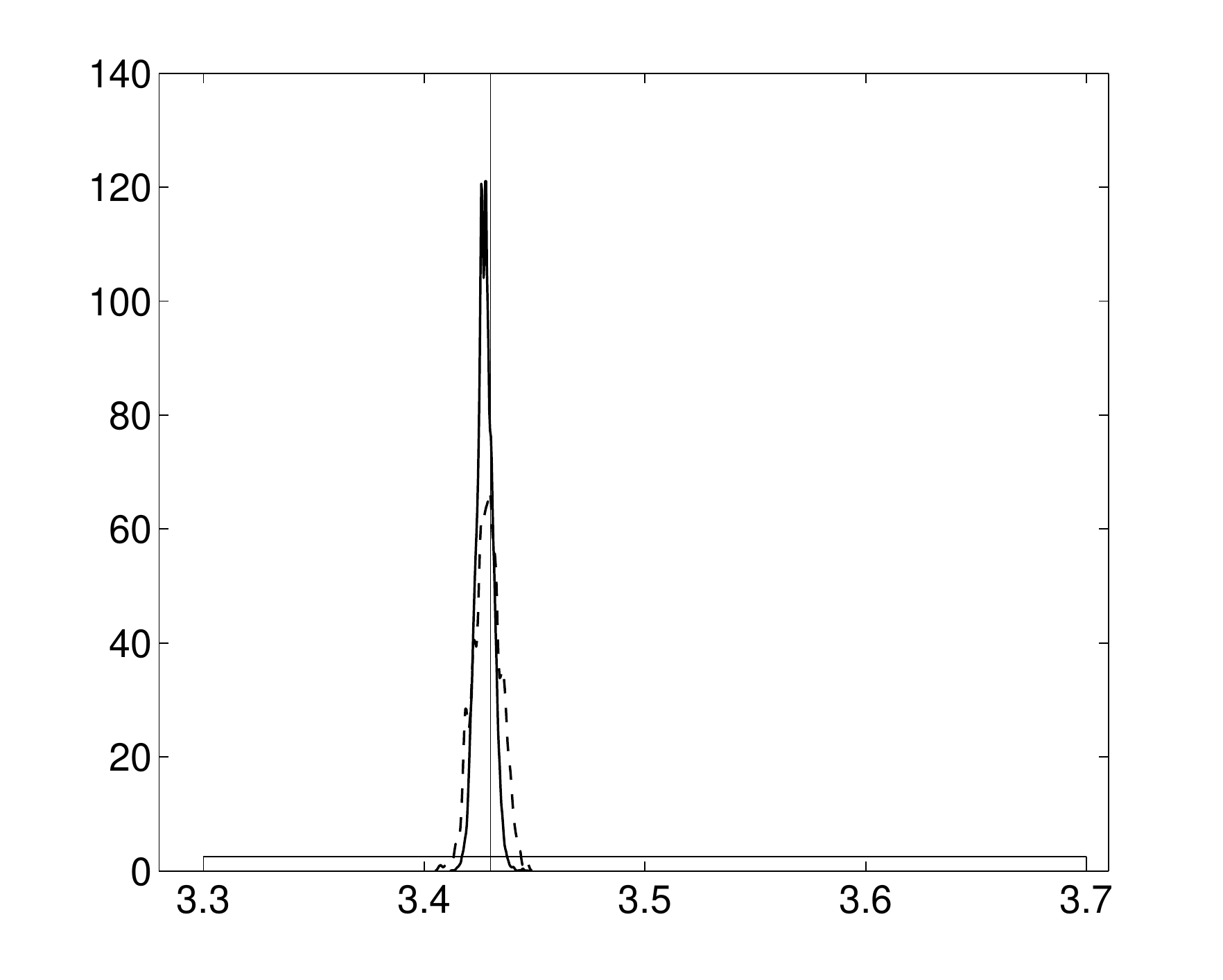}}\\
\subfigure[$\log \sigma_1$]{\includegraphics[scale=.35]{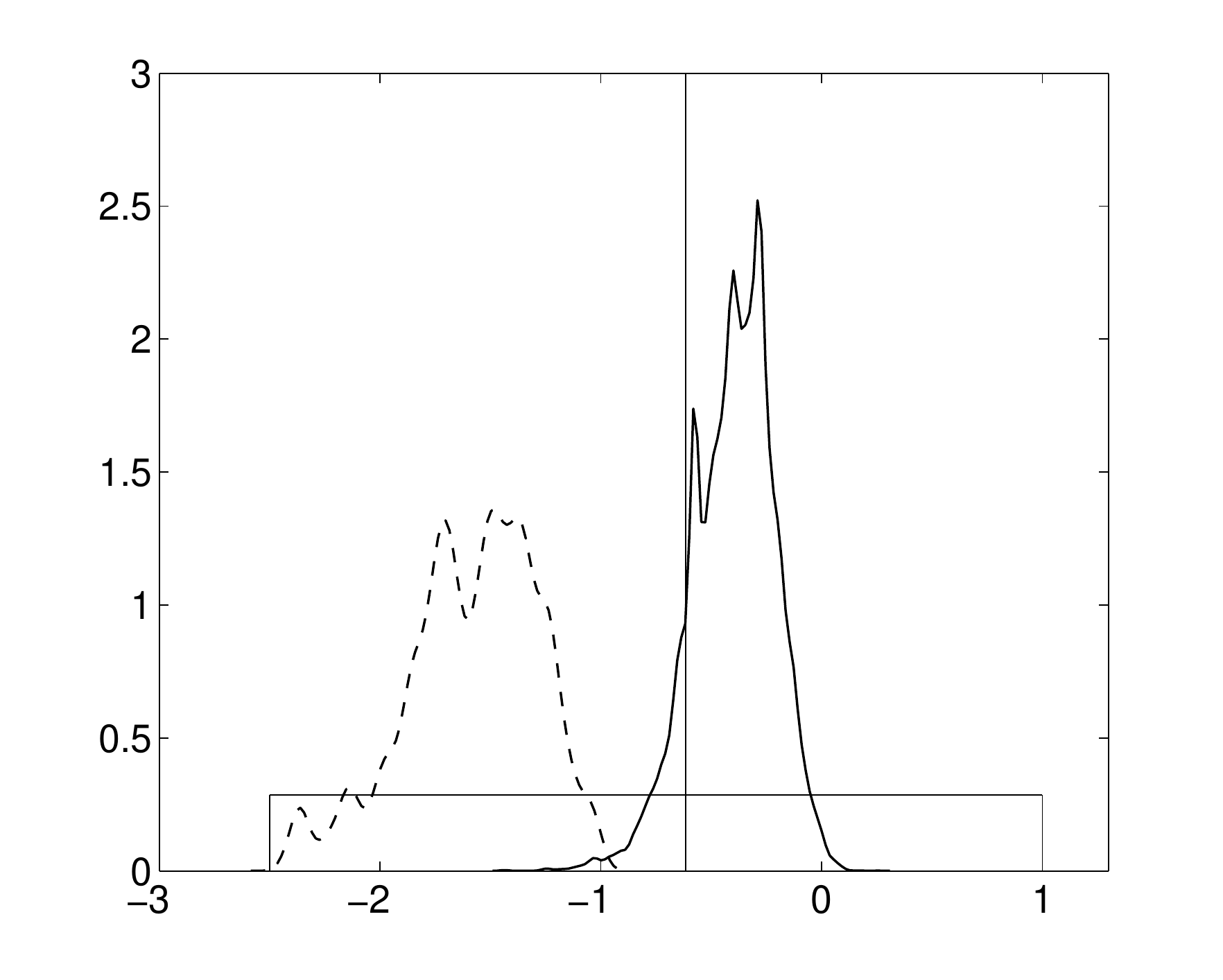}}
\subfigure[$\log \sigma_2$]{\includegraphics[scale=.35]{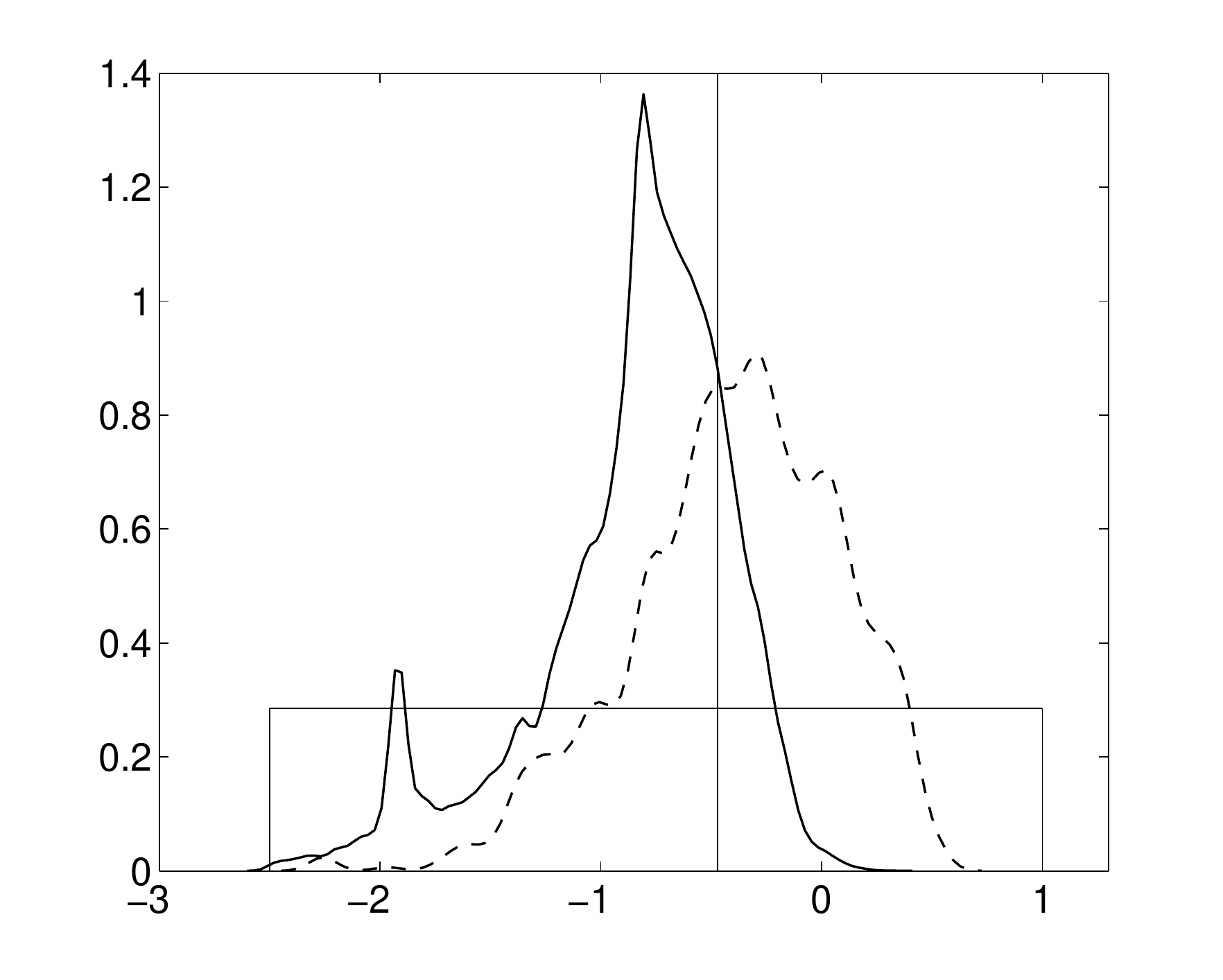}}

\caption{\footnotesize{Inference from simulated data when $n=355$: posterior
    marginal densities (by kernel smoothing approximation) for
    ABC-MCMC (dashed lines), particle MCMC (solid lines) and uniform
    priors. Vertical lines mark true parameter values. See the main text for further results.}}\label{fig:posteriors-pmcmc-abc}
\end{figure}

Figure \ref{fig:posteriors-pmcmc-abc} reports the estimated marginal
posterior densities from the particle MCMC and ABC-MCMC methods. The
``static'' features of the model represented by ${\bm \psi} =
(\mu_1,\mu_2,\sigma_1,\sigma_2,\alpha)$ seem to be overall well
captured by both inferential procedures. The modes $\mu_1$ and $\mu_2$
and the mixture parameter $\alpha$ can easily be identified, while the
variance parameters $\gamma$, $\sigma_1$, and $\sigma_2$ are somewhat
harder to identify and ABC appears to fail for $\sigma_1$ (this parameter is better estimated when using a larger sample size, see below). Regarding the correlation parameters $\theta$ and $\kappa$,
less information is available from the data and the posteriors do not
dominate strongly over the prior. In particular, $\kappa$ can hardly
be identified, which is likely due to the ``thinning'' of the data
leaving little information on the short scale correlation (recall that
$\kappa$ is the correlation parameter of the measurement error
process). This is confirmed by further results below as well as in section \ref{sec:real-application}, where several levels of subsampling are considered and the identification of $\kappa$ improves for a smaller $q$. We conclude that in this preliminary analysis ABC has shown an overall
satisfactory performance. We now produce further results using a larger sample size $n=n'=1,380$ while maintaining all other settings unchanged to obtain the following posterior inference (means and 95\% posterior intervals for each parameter), $\log\theta$:   $-6.275$ $[-6.950,-5.624]$,  $\log\kappa$:  $-0.538$ $[-1.407, 0.219]$, $\log\gamma$: $-0.027$ $[-0.540, 0.263]$,  $\log\mu_1$:  $3.24$ $[3.22,    3.25]$, $\log\mu_2$: $3.43$ $[3.41,3.45]$, $\log\sigma_1$: $-1.017$ $[-2.362,0.083]$,
$\log\sigma_2$:  $-0.854$ $[-2.240,0.419]$,  $\log\alpha$: $-0.697$ $[-1.159,-0.396]$. Clearly the estimation of $\kappa$ and $\sigma_1$ has improved. Unfortunately we cannot perform a comparison with particle MCMC when $n=1,380$ as this would require about 260 hrs of computation. Finally we check for possible improvements when using $n=1,380$ together with a larger set of percentiles in our vector of summary statistics (in addition to the usual autocorrelation values): we consider nine percentiles instead of six, i.e. the 10th, 20th,...,90th empirical percentiles. We select $\delta^*=0.4$ and obtain the following posterior inference: $\log\theta$:   $ -6.247$ $[ -6.734,   -5.706]$,  $\log\kappa$:  $-0.534$ $[ -1.448,0.220]$, $\log\gamma$: $0.003$ $[-0.371,0.225]$,  $\log\mu_1$:  $3.24$ $[3.22,3.25]$, $\log\mu_2$: $  3.43$ $[ 3.41,3.45]$, $\log\sigma_1$: $ -1.025$ $[ -2.419,    0.095]$,
$\log\sigma_2$:  $-0.877$ $[ -2.402,  0.321]$,  $\log\alpha$: $-0.627$ $[ -0.959,  -0.398]$. No striking difference emerges in comparison with the previous analysis, therefore we prefer to use only six percentiles, as the larger the size $d_s$ of $S(\cdot)$, when compared to $\mathrm{dim}(\theta)$, the larger the Monte Carlo error (Lemma 1 in \citep{fearnhead-prangle(2011)}).

%\end{document}

A striking difference between ABC and particle MCMC lies in the computational cost: for the case $n=355$ a cycle of 1,000 iterations of particle
MCMC is completed in 1,210 sec whereas for ABC-MCMC it requires only
6.5 sec. This makes it difficult to overlook an approximate inferential method such as ABC-MCMC.
Of course the price to be paid is the difficulty in tuning ABC
algorithms and most importantly choose the summary statistics. The
choice of kernel $K(\cdot)$ and tolerance $\delta$ is not
particularly challenging. However particle MCMC methods require not so much
tuning (once efficient proposal functions are constructed, and this is not an easy task in general) and return draws exactly from the posterior. Important examples of successful application of
ABC are e.g. \cite{barthelme-chopin} using expectation-propagation and
\cite{toni2009approximate} using SMC within ABC.

\section{Application: A protein folding problem}\label{sec:real-application}
\begin{figure}
\centering
\includegraphics[height=5cm,width=14cm]{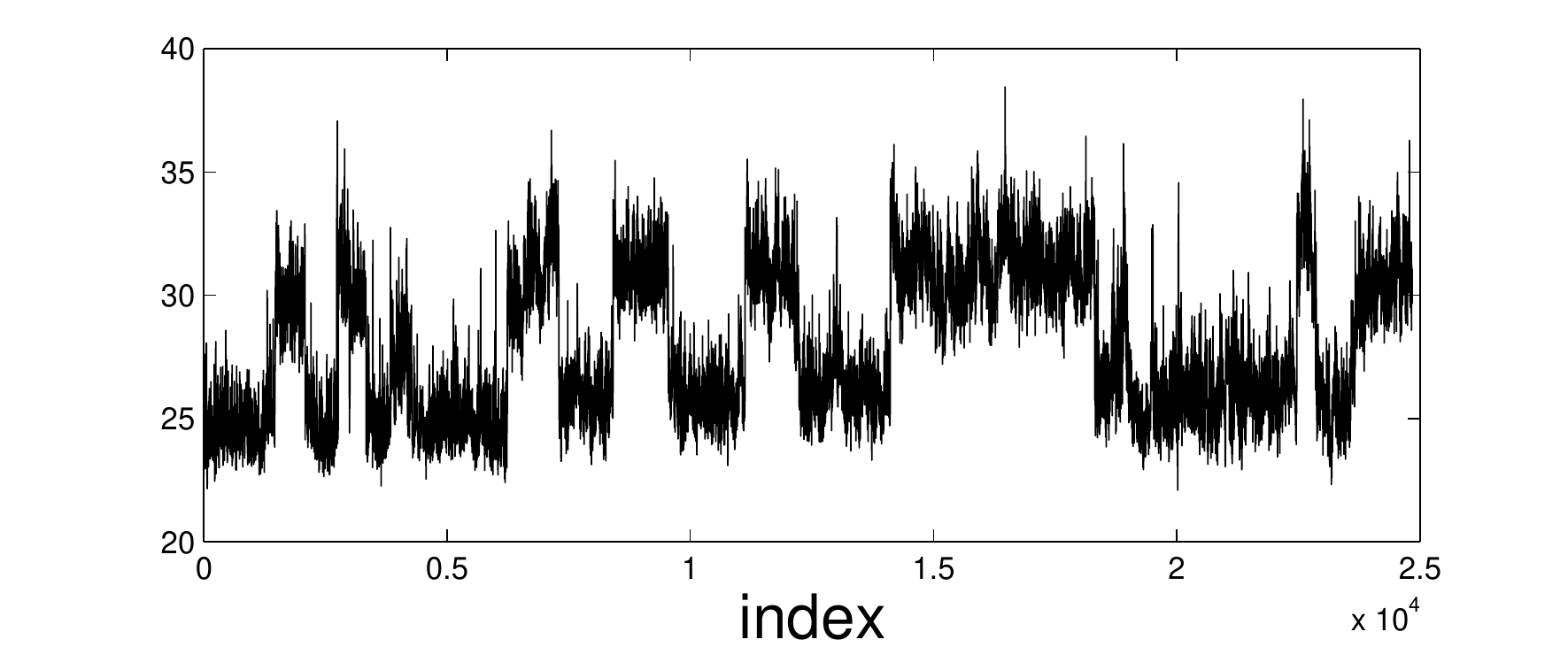} 
\includegraphics[height=5cm,width=9cm]{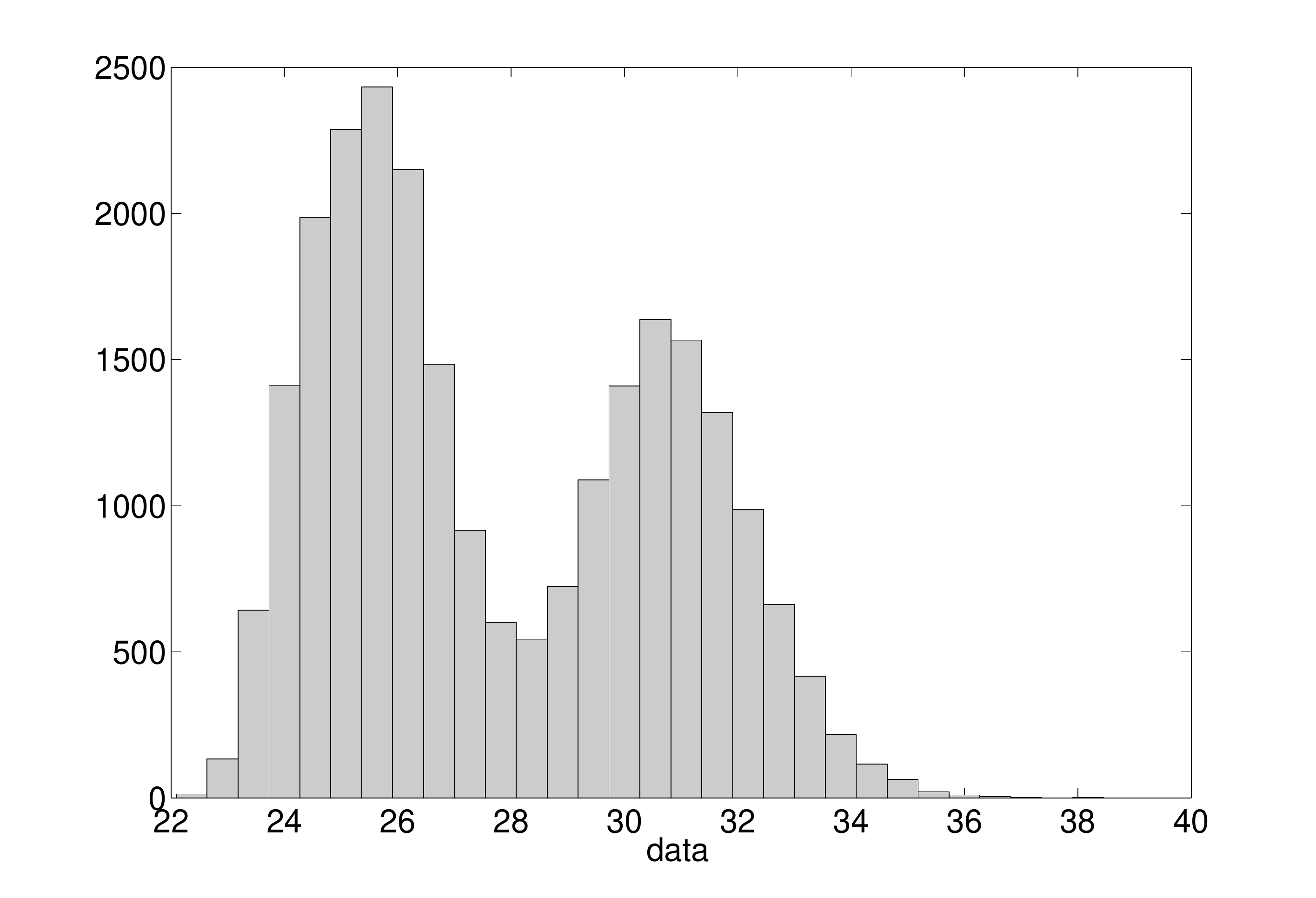} 
\caption{Sample path and sample histogram of the so-called $L$-projection of the small Trp-zipper protein. The distribution of the data reflects the two states of the protein.}
\label{fig:data}
\end{figure}

Proteins are synthesized in the cell on ribosomes as linear,
unstructured polymers that self-assemble into specific and functional
three-dimensional structures. This self-assembly process, called
\textit{protein folding}, is the last and crucial step in the
transformation of genetic information, encoded in DNA, into functional
protein molecules. Because of its biological importance, the
understanding of protein folding has received enormous interest both
in experiments, theory and simulations \citep{wolynes2012chemical}.
For reasons of simplification and tractability, the dynamics of a
protein are often modelled as diffusions along a single
\textit{reaction coordinate}, that is one-dimensional diffusion models
are considered to model a projection of the actual dynamics in
high-dimensional space, see \cite{sow:96,das:06} and references therein.
In our case study we consider the so-called L-reaction coordinate of 
the small Trp-zipper protein with $n=24,842$ observations taken at a
sampling frequency of $\Delta^{-1}=$1/nsec. 
The high-dimensional dynamics of the protein were simulated 
from the Monte Carlo algorithm of \cite{sb:12} using the PHAISTOS
software package \cite{wb:13}. Alongside the L-reaction coordinate was computed.
The sample path of the reaction coordinate, Figure \ref{fig:data}, clearly 
reflects the random switching of the protein between the folded (lower
mode) and unfolded (upper mode) state.

In a preliminary analysis, \cite{fs:14} found that these data were not
well fitted by any Markovian model, but that the diffusion observed
with measurement error model \eqref{eq:state-space} gave a good fit
both on the short and on the long time scale.

To estimate the parameter ${\bm \eta}$ from the protein data we apply the
ABC algorithm \ref{alg:lfmcmc-earlyrej}. The priors and the overall
setup is the same as in the simulation study but this time we use
subsampling in the simulations within algorithm \ref{alg:lfmcmc-earlyrej}.
In fact we perform three studies where a different value for the subsample size $n'$ (hence a different $q$) is considered in each case: in the first study trajectories are simulated in correspondence of every
$q=30$'th observation of the full data, so that $n'=829$ and the ${\bm x}_{sim}$'s (and ${\bm z}_{sim}$'s)
are simulated at times $\{t_0,t_{30},t_{60},...,t_{n-30},t_n\}=\{1,31,61,...,24841\}$. Similarly in the other two studies we choose $q=15$ ($n'=1,657$) and $q=7$ ($n'=3,549$) respectively, and corresponding time grids.
The algorithm assumes the initial state for $\{X_t\}$ to be a
known constant $x_0$. Since $U\approx 0$ we have
$x_0\approx\tau^{-1}(z_0)$. The initial observation $z_0=23.248$ corresponds
to the empirical $0.0072$-quantile of the data. Hence, in the three studies we set $x_0
=\Phi^{-1}(0.0072)= -2.45$. 

\begin{figure}
\centering
\subfigure[Subfigure 1 list of figures text][$\log \theta$]{\includegraphics[ height=4cm, width=6cm]{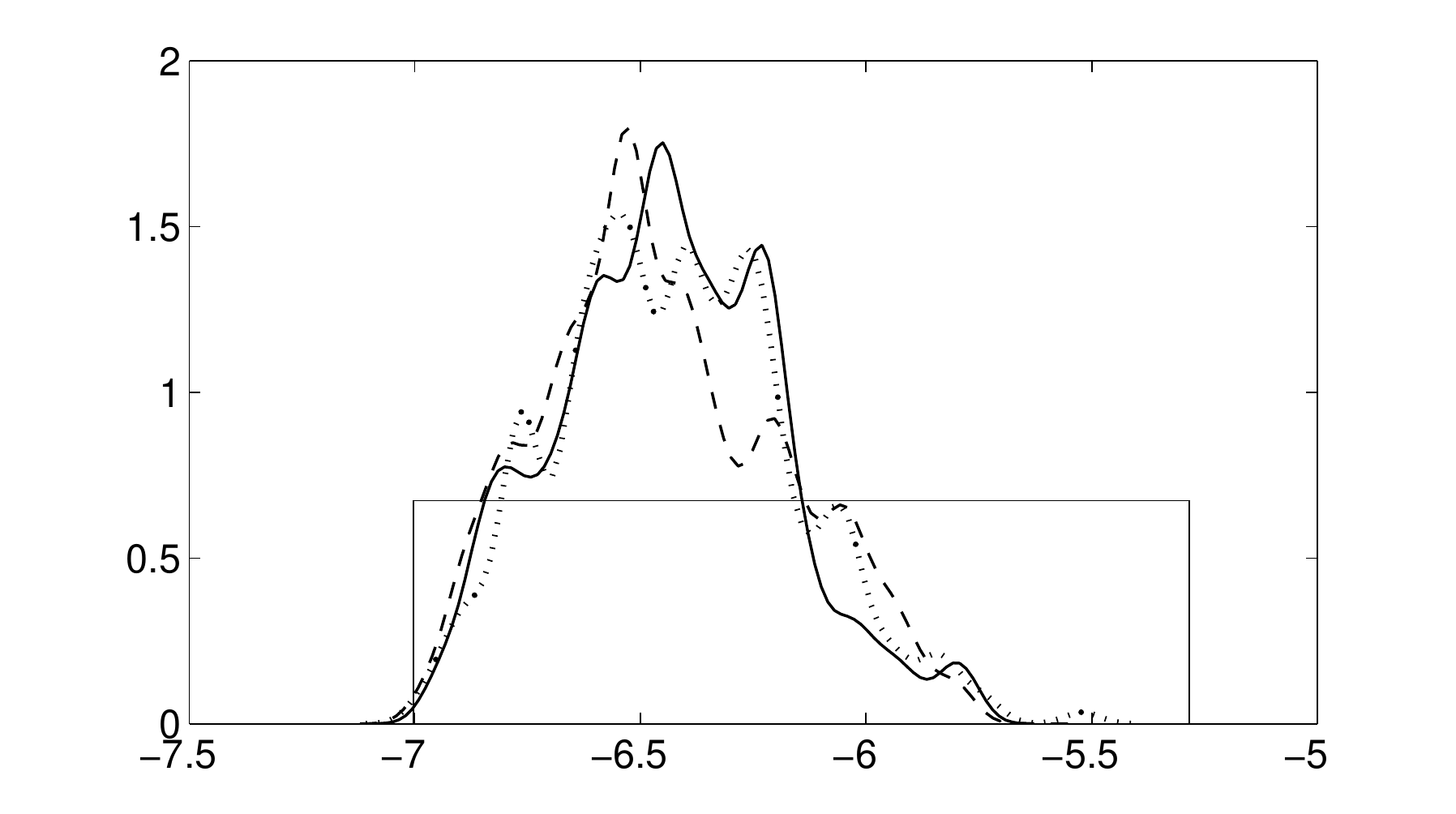}}
\subfigure[Subfigure 1 list of figures text][$\log \kappa$]{\includegraphics[ height=4cm, width=6cm]{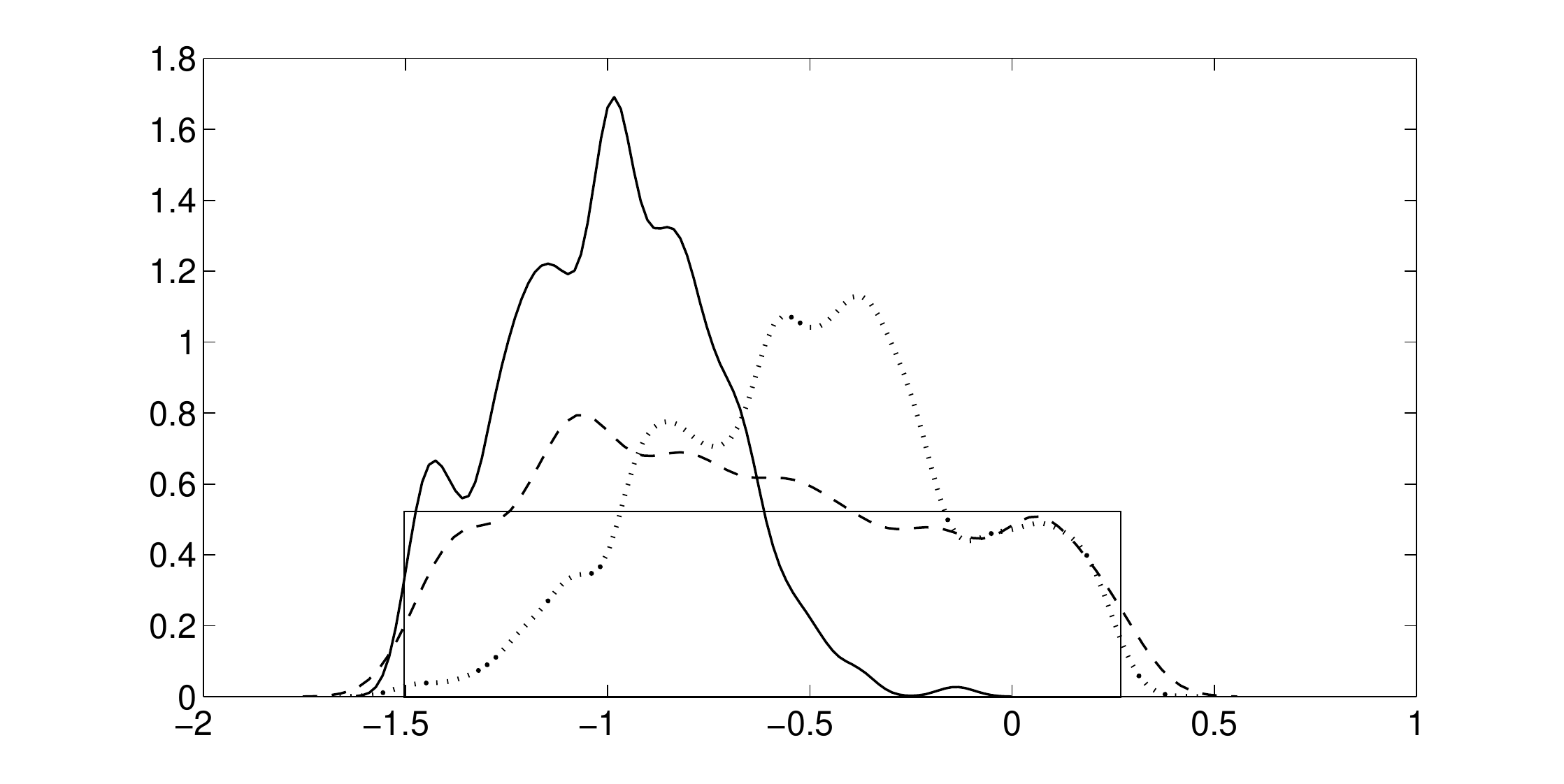}}\\
\subfigure[Subfigure 1 list of figures text][$\log \gamma$]
{\includegraphics[height=4cm, width=6cm]{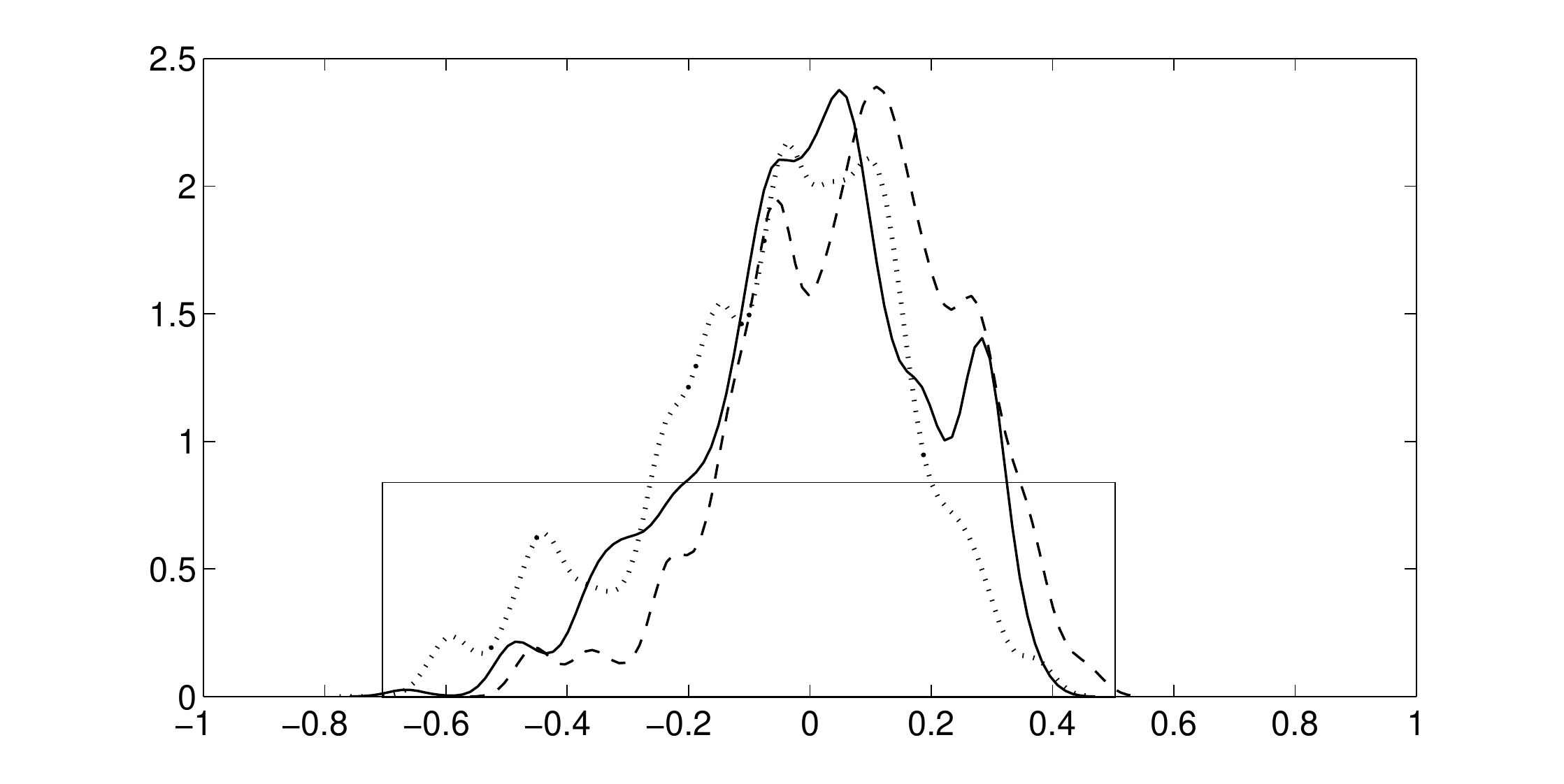}}
\subfigure[Subfigure 1 list of figures text][$\log \alpha$]{\includegraphics[height=4cm, width=6cm]{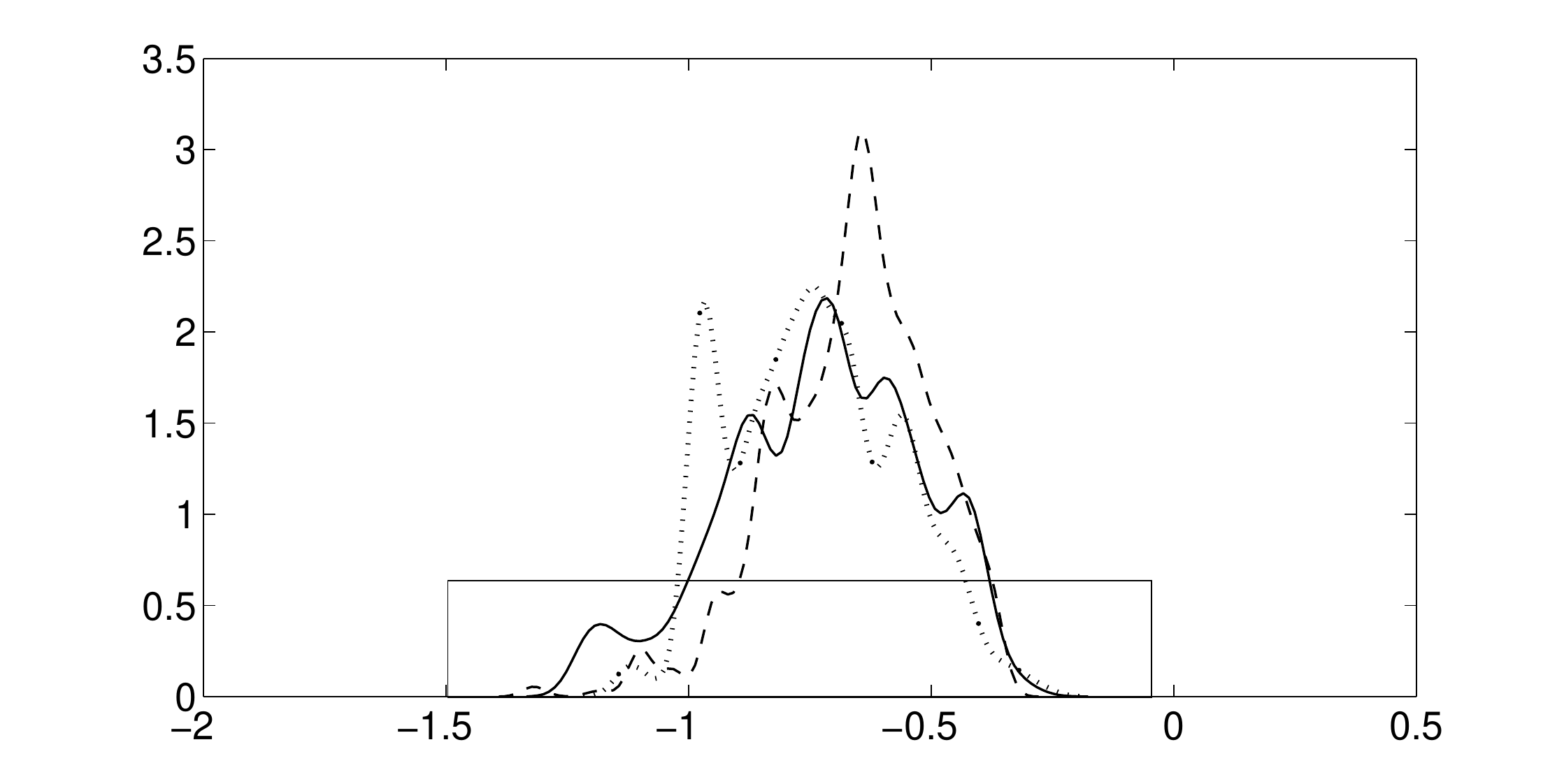}}\\
\subfigure[Subfigure 1 list of figures text][$\log \mu_1$]{\includegraphics[height=4cm, width=6cm]{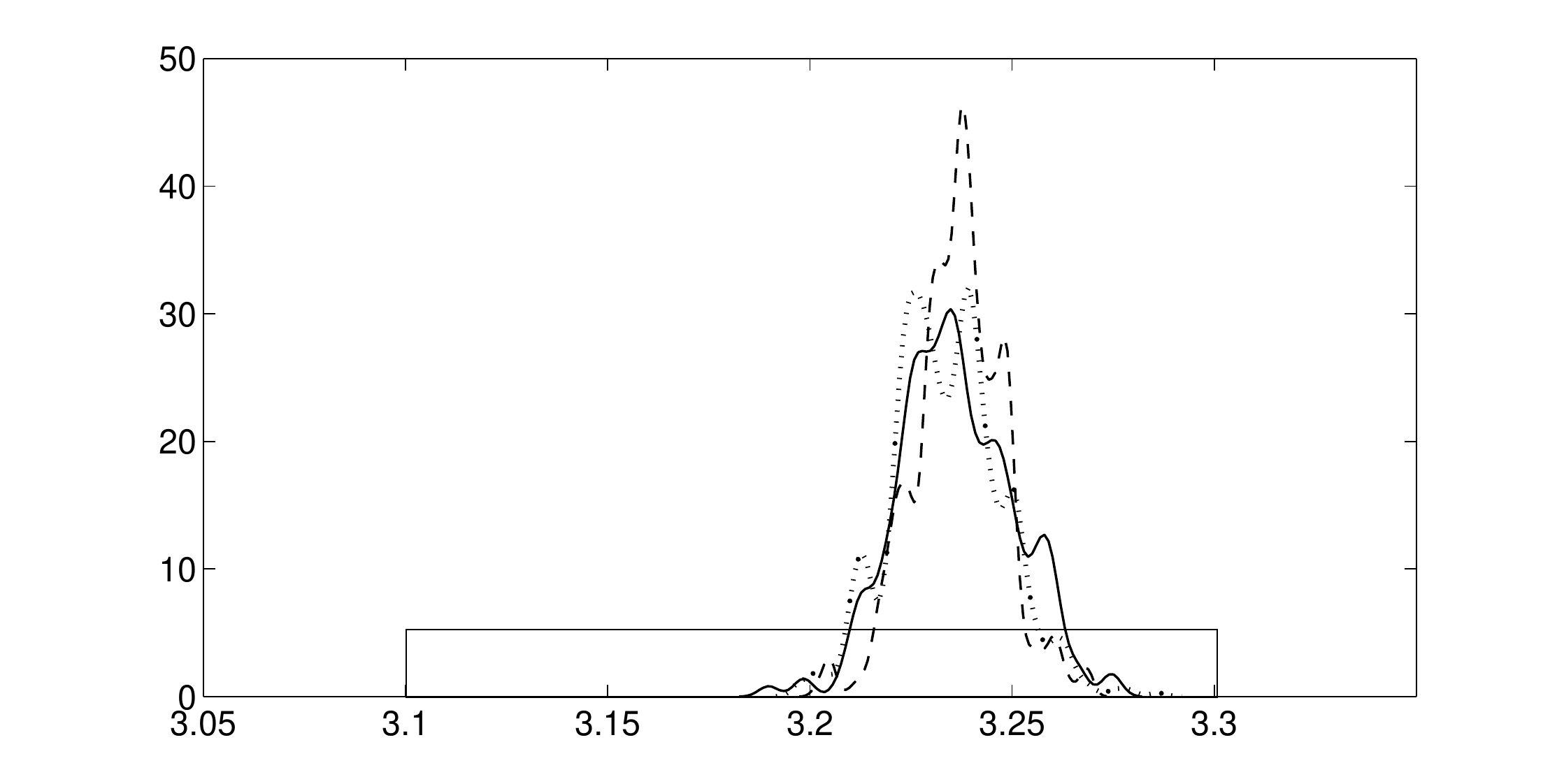}}
\subfigure[Subfigure 1 list of figures text][$\log \mu_2$]{\includegraphics[height=4cm, width=6cm]{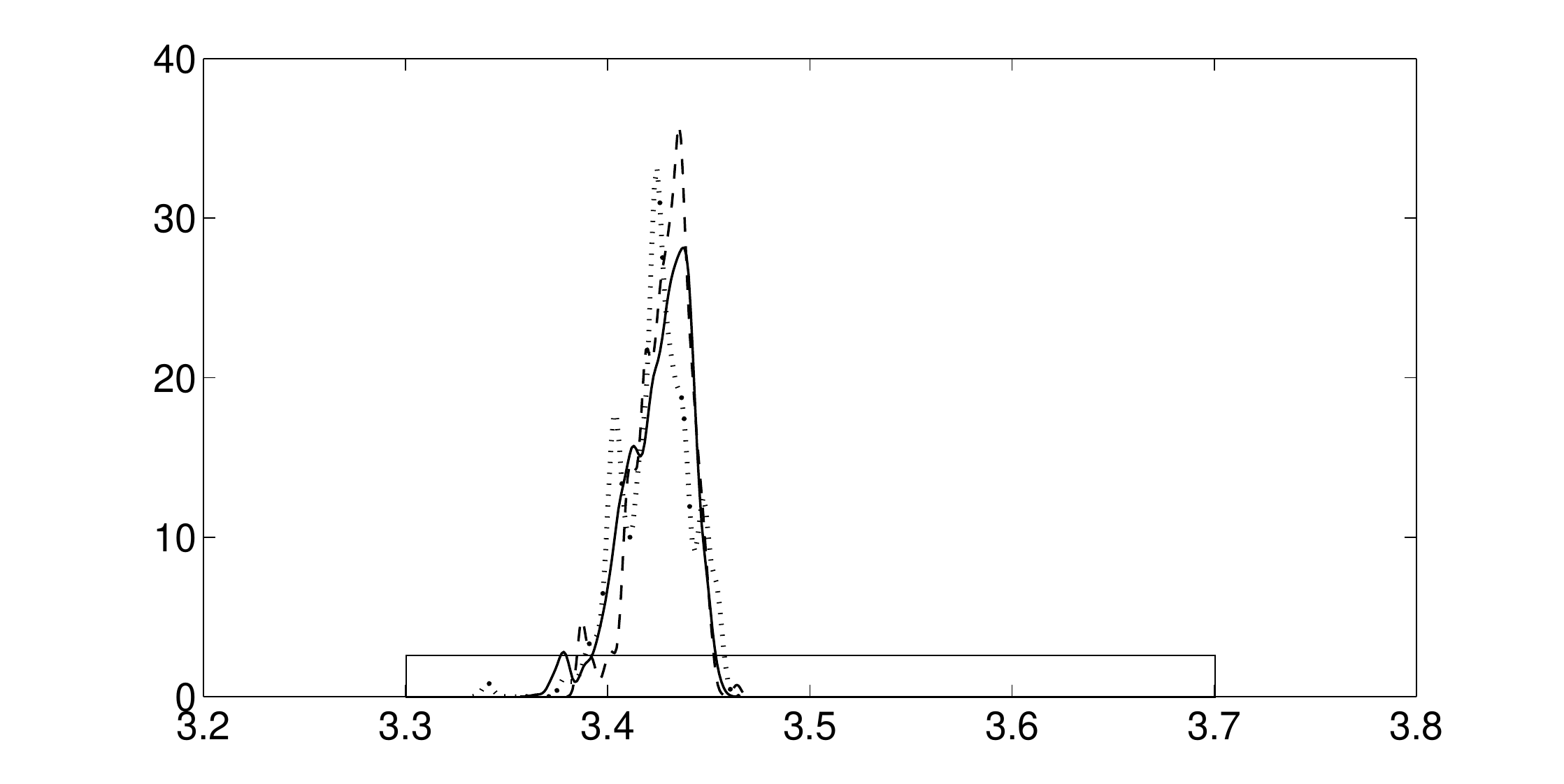}}\\
\subfigure[Subfigure 1 list of figures text][$\log \sigma_1$]{\includegraphics[height=4cm, width=6cm]{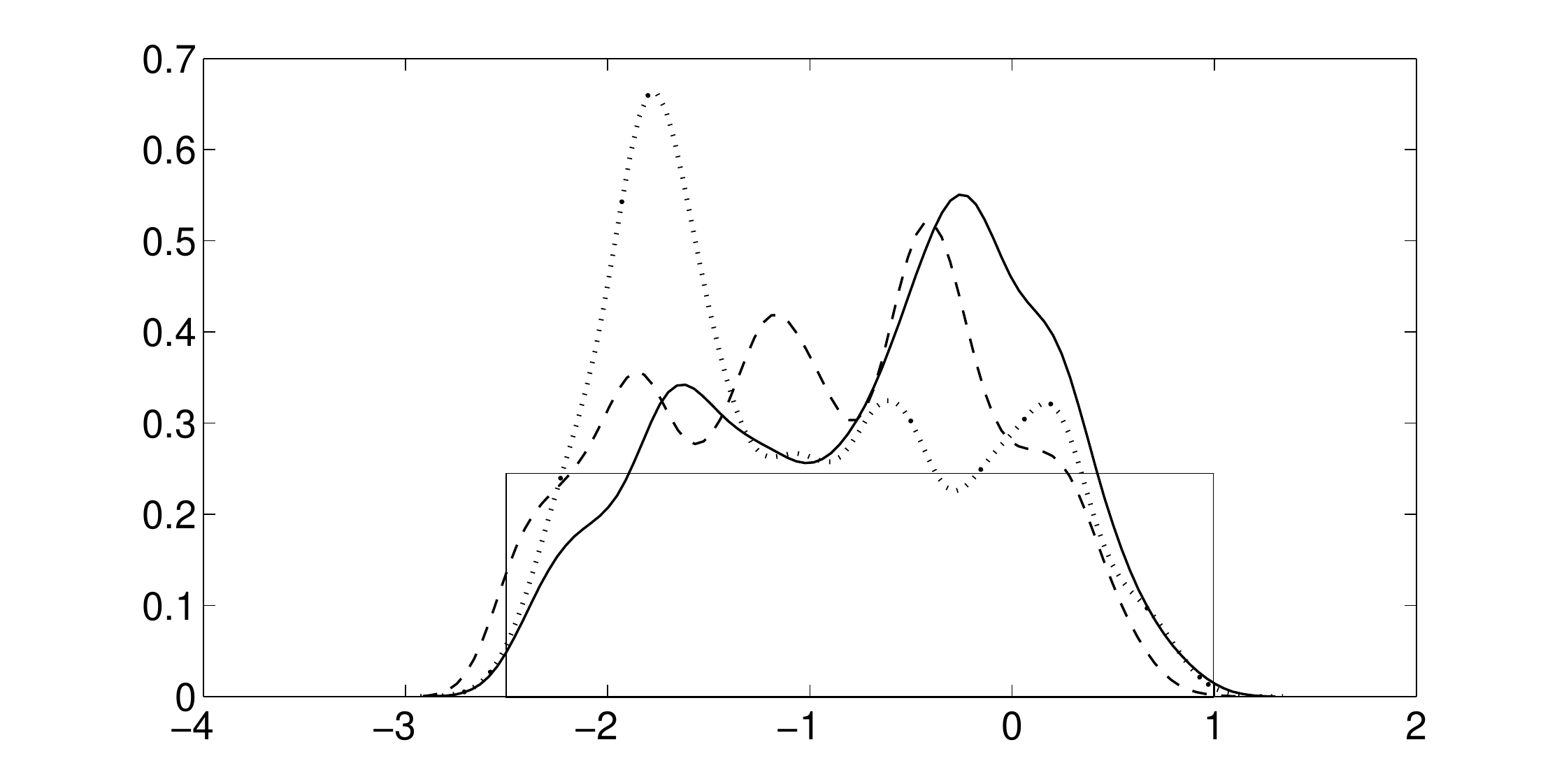}}
\subfigure[Subfigure 1 list of figures text][$\log \sigma_2$]{\includegraphics[height=4cm, width=6cm]{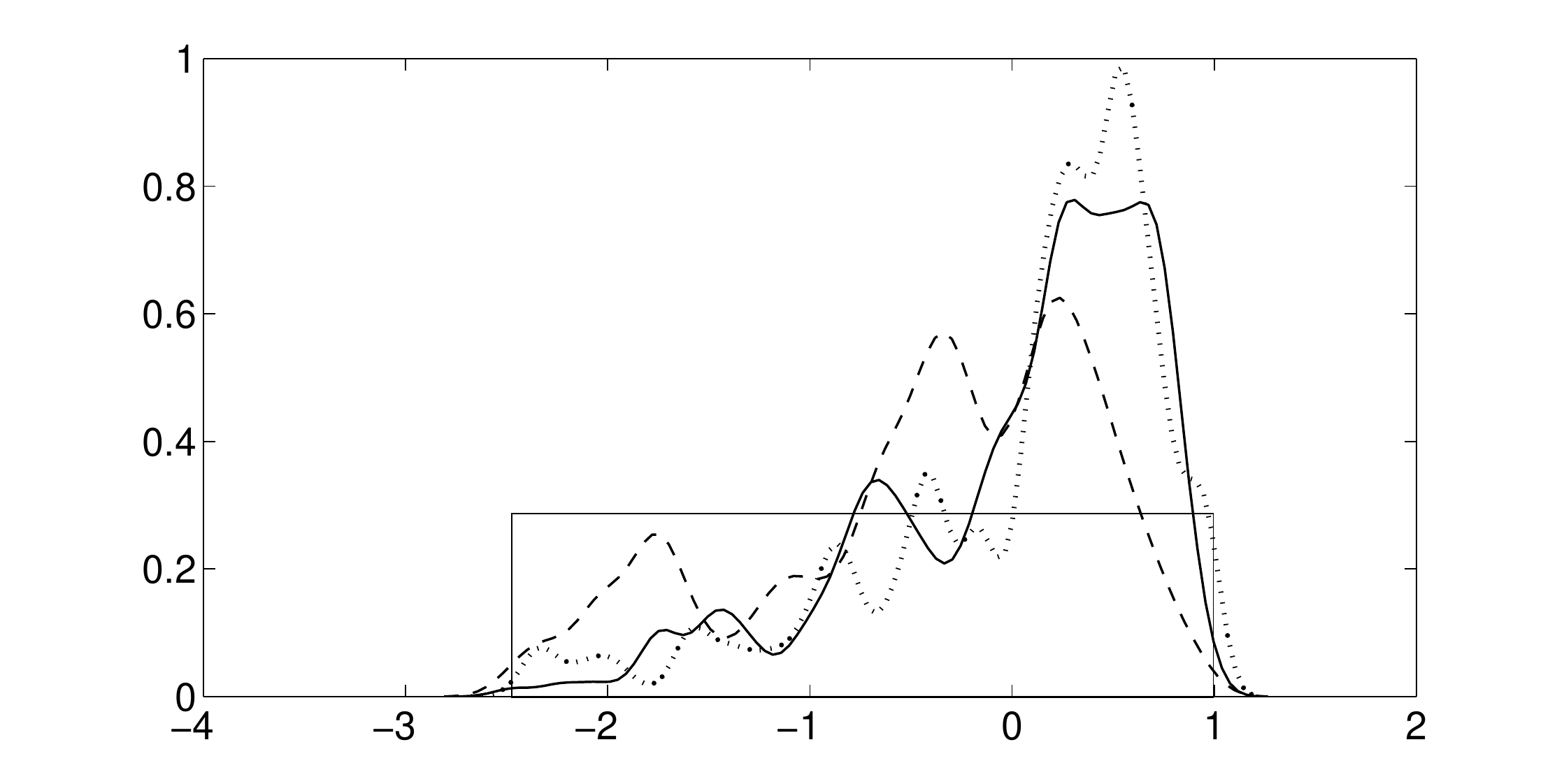}}
\caption{ABC inference from protein folding data: approximated
  marginal posteriors for subsamples having size $n'=829$ (dashed line), $n'=1657$ (dotted), $n'=3549$ (solid) and uniform priors.}\label{fig:prior-posterior-compare}
\end{figure}

We now start discussing the experiment with $n'=829$. As anticipated in section \ref{sec:early-rej-abc}, 
we take values of the autocorrelation function as summary
statistics, namely the ones at lags $(60,300,600,1200,1800,2100)$ for observed data
${\bm z}$ and at lags $(2,10,20,40,60,70)$ for ${\bm z}_{sim}$. Additionally we use the
15th, 30th, 45th, 60th, 75th and 90th empirical percentiles of both observed and simulated data as summary
statistics. Thus ${\bm S}_n(\cdot)$ and ${\bm S}_{n'}(\cdot)$ have length
$d_s=12$. Finally we set $\delta_{start}=0.7$, $\delta_{max}=0.9$, $\delta_{minmax}=0.65$. 
Algorithm \ref{alg:lfmcmc-earlyrej} was run for
$R=2\times 10^6$ iterations, thinning every 10th draw and obtaining an
average acceptance rate of about 1\%. The simulation was completed in about 6.3 hrs when $n'=829$. 
Same as in section \ref{sec:simulation-study} we observed how the posterior means of the
ABC output change for varying values of $\delta$ and decided to
filter-out draws corresponding to $\delta>\delta^*=0.45$. Results from the
remaining 16,000 draws are shown in Table \ref{tab:estimates-realdata} and
Figure \ref{fig:prior-posterior-compare}. Same as before parameters $\kappa$ and $\sigma_1$
are quite uncertain, while the other parameters appears to be well
identified from the data. In particular, we expect $\kappa$ to be better
identified when increasing the size of the subsample, and this is confirmed in the other two studies. When experimenting with $n'=1,657$ and $n'=3,549$ we keep the same simulation settings as detailed above, including the choice $\delta^*=0.45$, and in the first case results were returned in 12.5 hrs and in 27.5 hrs in the second case. Results are compared in Table \ref{tab:estimates-realdata} and
Figure \ref{fig:prior-posterior-compare}. As expected, for increasing $n'$ we note a markedly different approximated posterior for $\log \kappa$, because such parameter enters the autocorrelation function for the Ornstein-Uhlenbeck model $\{U_t\}$ and therefore a different subsampling has an effect on the autocorrelation function, hence an effect on $\kappa$. It is reassuring not to spot serious differences in the inference for the remaining parameters (except for $\sigma_1$), this implying that the information explained by our model \eqref{eq:state-space} and contained in our summary statistics is preserved for different levels of subsampling and that a ``harder'' subsampling ($q=30$) does not seem to have a major influence on overall results.

\begin{table}
\caption{Protein folding data experiment: posterior means from the \textit{filtered} ABC-MCMC output and 95\% posterior intervals for the cases $n'=829$ (first line of each estimated parameter), $n'=1657$ (second line) and $n'=3549$ (third line).}
\begin{tabular}{lcl}
\hline
\hline
{} & ABC inference \\ 
\hline
$\log\theta$   & --6.448 &[--6.646,--5.909]   \\
{} & --6.421 & [--6.899,--5.847]\\
{} & --6.438 & [--6.863,--5.891]\\
$\log\kappa$   & --0.649 & [--1.054,0.246]   \\
{} & --0.492 & [--1.202,0.185]\\
{} & -0.996 & [--1.468,--0.522]\\
$\log\gamma$   & 0.070 & [--0.052,0.378]   \\
{} & -0.055 & [--0.491,0.279]\\
{} & 0.005 & [--0.385,0.310]\\
$\log\mu_1$    & 3.24  &[3.23,3.26]   \\ 
{} & 3.23 & [3.21,3.26]\\
{} & 3.24 & [3.21,3.26]\\
$\log\mu_2$    & 3.43 & [3.42,3.45]  \\ 
{} & 3.42 &  [3.39,3.45]\\
{} & 3.43 & [3.38,3.45]\\
$\log\sigma_1$ & --0.962 &  [--1.665,0.364]  \\ 
{} & --1.044 & [--2.276,0.601]\\
{} & --0.719 & [--2.269,0.546]\\
$\log\sigma_2$ & --0.418 &  [--0.862,0.765]  \\ 
{} & 0.039 & [--2.074,0.957]\\
{} & 0.006 & [--1.752,0.864]\\
$\log\alpha$   & --0.663 & [ --0.766,--0.383]    \\ 
{} & --0.741 & [--0.996,--0.420]\\
{} & --0.725 & [--1.188,--0.399]\\
\hline
\end{tabular}
\label{tab:estimates-realdata}
\end{table}

\begin{figure}
\centering
\includegraphics[scale=0.9]{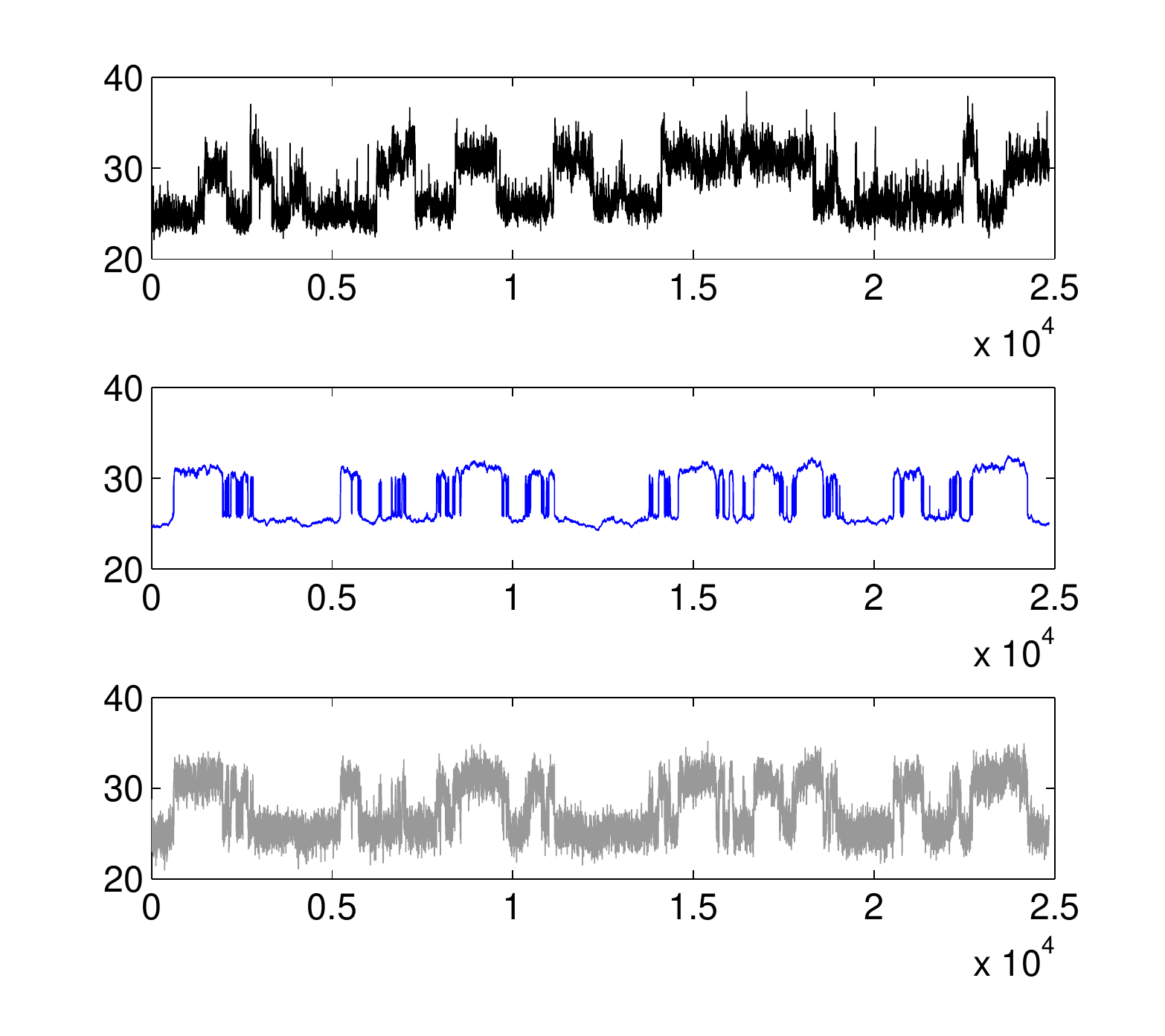}
\caption{Comparison of protein folding data (top) with estimated
  processes $\tau(X_t)$ (middle) and  $Z_t$
  (bottom) based on model \ref{eq:state-space} with parameters
  set to the estimated posterior means obtained for the case $n'=829$.}\label{fig:realdata-fit}
\end{figure}

As a final informal check of our result we generated a time
series of size $n=24,842$ from model \eqref{eq:state-space} using parameters equal to the
posterior means obtained for the case $n'=829$. The sample path
is compared to that of observed data in Figure
\ref{fig:realdata-fit}. Although not perfect, the parameter estimates
seem to capture the overall features in the data including timely
switching between the two states. Corresponding trajectories for the case $n'=3549$ do not result in noticeable differences and are thus not reported. 

\section{Discussion}
We have considered a complex stochastic dynamical model
in form of a nonlinear diffusion observed
with measurement error having a bimodal marginal structure with
correlated error terms. The model has applications to a
protein-folding problem where the data has size $n\approx 25,000$. 
Both the model and the size of data pose several problems both from a
computational and a methodological point of view: (i) data analysed
with the considered model are not conditionally independent given the
latent state. This prevents the use of methods for state
space models. (ii) The size of data prohibits the use of suitable but
computer-intensive methods based on sequential Monte Carlo and
likelihood-free  Markov chain Monte Carlo (MCMC) algorithms. We
proposed to conduct inference using approximated Bayesian computation
(ABC) as a reasonable compromise between likelihood based inference
and computational feasibility. An important feature of ABC is
the ability to exploit the information carried by the data by means of
summary statistics. We found that in our case ABC enables inference in
a large data context by use of ``subsampling'', that is while the
entire dataset was used for inference, shorter trajectories, i.e.\
subsamples, were simulated within an ABC MCMC algorithm. Avoiding
expensive simulations of latent trajectories having the same size as the
available data is a major improvement in terms of time
consumption and inferential results were encouraging. 
In fact the several levels of subsampling we investigated seem to affect only a small number of quantities in our model (specifically $\kappa$ and $\sigma_1$). This means that the speed we gain by simulating shorter trajectories does not translate in a significant loss of information, which is one of the advantages of using ABC, meaning that when available information is exploited via appropriate summaries (even if not sufficient statistics) satisfactory results can be obtained at a fraction of the cost corresponding to using the full data.
Thus in the present
case study the ABC method offered a valid alternative to exact but computationally expensive methodologies. 

Other successful applications of subsampling can be found
in \cite{ahn-et-al,korattikara2013austerity} and it should
be noted that it makes sense to consider subsampling for our specific
application where dynamics follow a characteristic stationary
pattern. In other applications, using subsampling may or may not be
appropriate. Relevant and
crucial comments on an early version of the present work have been
raised in Christian P. Robert's blog\footnote{\url{http://xianblog.wordpress.com/2013/10/17/accelerated-abc/}}:
one concern was the increased variability of the summary statistics
when evaluated on a subsample. Given that we subsample dynamics having a fairly regular
pattern, we expect that subsampling may lead to more variable results but
not add any substantial bias. More in detail, both the empirical quantiles, empirical moments, and the empirical joint moments entering the summary statistics are M-estimators. Hence both the full sample and subsampled summary statistics are $\sqrt{n}$--consistent and asymptotically normal estimators of the true quantile/correlation-vector under suitable regularity conditions (see e.g. \cite{newey1990semiparametric}) for large sample sizes and fixed subsampling level $q$. Under the true data generating measure the difference between the full data summary statistics and subsampled statistics generated independently hereof is thus approximately multivariate normal, with zero mean and a covariance matrix which could be derived from the asymptotic expansions of the estimators. This suggests that the inverse of the covariance for the difference would be an optimal weight in the distance measure and considered as matrix $\mathbf{A}$ into \eqref{eq:unikernel}. However, the covariance matrix/optimal weight is in practice unknown as it depends on the true parameter. Ad hoc selections of this matrix are expected to produce asymptotically unbiased but suboptimal estimates. Note that subsampling in itself reduces efficiency as the asymptotic variance of the summary statistics is multiplied by a factor $q$. Giving a more formal account of the asymptotic properties of our estimators is technical and beyond the scope of the present paper.
 
\section*{Acknowledgements} 
We are grateful to Sandro Bottaro and Jesper Ferkinghoff-Borg, Elektro
DTU, for supplying the data for the case study. We thank an anonymous reviewer for providing useful suggestions that improved the present work.

\section*{Funding} 
Umberto Picchini and Julie Forman research is partly funded by a grant from the Swedish Research Council (VR grant 2013-5167).

\bibliography{newbiblio}

\begin{thebibliography}{41}
\providecommand{\natexlab}[1]{#1}
\providecommand{\url}[1]{\texttt{#1}}
\expandafter\ifx\csname urlstyle\endcsname\relax
  \providecommand{\doi}[1]{doi: #1}\else
  \providecommand{\doi}{doi: \begingroup \urlstyle{rm}\Url}\fi

\bibitem[Ahn et~al.(2012)Ahn, Korattikara, and Welling]{ahn-et-al}
S.~Ahn, A.~Korattikara, and M.~Welling.
\newblock Bayesian posterior sampling via stochastic gradient {F}isher scoring.
\newblock In John Langford and Joelle Pineau, editors, \emph{Proceedings of the
  29th International Conference on Machine Learning}, pages 1591--1598, 2012.
\newblock \texttt{arXiv:1206.6380}.

\bibitem[A{\"{i}}t-Sahalia(1996)]{yas:96}
Y.~A{\"{i}}t-Sahalia.
\newblock Testing continuous-time models of the spot interest rate.
\newblock \emph{The Review of Financial Studies}, 9:\penalty0 385--426, 1996.

\bibitem[Andrieu et~al.(2010)Andrieu, Doucet, and
  Holenstein]{andrieu2010particle}
C.~Andrieu, A.~Doucet, and R.~Holenstein.
\newblock Particle {M}arkov chain {M}onte {C}arlo methods (with discussion).
\newblock \emph{Journal of the Royal Statistical Society: Series B},
  72\penalty0 (3):\penalty0 269--342, 2010.

\bibitem[Azencott et~al.(2013)Azencott, Arjun, Ankita, and Timofeyev]{Azencott}
R.~Azencott, B.~Arjun, J.~Ankita, and I.~Timofeyev.
\newblock Sub-sampling and parameter estimation for multiscale dynamics.
\newblock \emph{Communications in Mathematical Sciences}, 11:\penalty0
  939--970, 2013.

\bibitem[Barthelm\'{e} and Chopin(2014)]{barthelme-chopin}
S.~Barthelm\'{e} and N.~Chopin.
\newblock {Expectation propagation for likelihood-free inference}.
\newblock \emph{Journal of the American Statistical Association}, 109\penalty0
  (505):\penalty0 315--333, 2014.

\bibitem[Beaumont(2010)]{beaumont2010approximate}
M.A. Beaumont.
\newblock Approximate {B}ayesian {C}omputation in evolution and ecology.
\newblock \emph{Annual Review of Ecology, Evolution, and Systematics},
  41:\penalty0 379--406, 2010.

\bibitem[Best and Hummer(2010)]{bh:10}
R.~B. Best and G.~Hummer.
\newblock Coordinate-dependent diffusion in protein folding.
\newblock \emph{PNAS}, 107:\penalty0 1088--1093, 2010.

\bibitem[Bezanson et~al.(2012)Bezanson, Karpinskiy, Shah, and Edelman]{julia}
J.~Bezanson, S.~Karpinskiy, V.~B. Shah, and A.~Edelman.
\newblock Julia: A fast dynamic language for technical computing.
\newblock \texttt{arXiv:1209.5145v1}, 2012.

\bibitem[Boomsma et~al.(2013)]{wb:13}
W.~Boomsma et~al.
\newblock {PHAISTOS}: A framework for {M}arkov chain {M}onte {C}arlo simulation
  and inference of protein structure.
\newblock \emph{Journal of Computational Chemistry}, 34\penalty0 (19):\penalty0
  1697--1705, 2013.

\bibitem[Bortot et~al.(2007)Bortot, Coles, and Sisson]{bortot2007inference}
P.~Bortot, S.G. Coles, and S.~Sisson.
\newblock Inference for stereological extremes.
\newblock \emph{Journal of the American Statistical Association}, 102\penalty0
  (477):\penalty0 84--92, 2007.

\bibitem[Bottaro et~al.(2012)Bottaro, Boomsma, Johansson, Andreetta, Hamelryck,
  and Ferkinghoff-Borg]{sb:12}
S.~Bottaro, W.~E. Boomsma, K.~Johansson, C.~Andreetta, T.~Hamelryck, and
  J.~Ferkinghoff-Borg.
\newblock Subtle {M}onte {C}arlo updates in dense molecular systems.
\newblock \emph{Journal of Chemical Theory and Computation}, 8:\penalty0
  695--702, 2012.

\bibitem[Bret{\'o} et~al.(2009)Bret{\'o}, He, Ionides, and King]{breto2009time}
C.~Bret{\'o}, D.~He, E.~L. Ionides, and A.~A. King.
\newblock Time series analysis via mechanistic models.
\newblock \emph{The Annals of Applied Statistics}, 3\penalty0 (1):\penalty0
  319--348, 2009.

\bibitem[Crommelin and Vanden-Eijnden(2011)]{Eijnden}
D.~Crommelin and E.~Vanden-Eijnden.
\newblock Diffusion estimation from multiscale data by operator eigenpairs.
\newblock \emph{SIAM Multiscale Modeling and Simulation}, 9:\penalty0
  1588--1623, 2011.

\bibitem[Das et~al.(2006)Das, Moll, Stamati, Kavraki, and Clementi]{das:06}
P.~Das, M.~Moll, H.~Stamati, L.~E. Kavraki, and C.~Clementi.
\newblock Low-dimensional, free-energy landscapes of protein-folding reactions
  by nonlinear dimensionality reduction.
\newblock \emph{PNAS}, 103:\penalty0 9885--9890, 2006.

\bibitem[Doucet et~al.(2001)Doucet, De~Freitas, and
  Gordon]{doucet2001sequential}
A.~Doucet, N.~De~Freitas, and N.~Gordon.
\newblock \emph{Sequential {M}onte {C}arlo methods in practice}.
\newblock Springer New York, 2001.

\bibitem[Drovandi(2014)]{drovandi(2013)}
C.~Drovandi.
\newblock Pseudo-marginal algorithms with multiple {CPU}s.
\newblock Queensland University of Technology, available at
  \url{http://eprints.qut.edu.au/61505/}, 2014.

\bibitem[Fearnhead and Prangle(2012)]{fearnhead-prangle(2011)}
P.~Fearnhead and D.~Prangle.
\newblock Constructing summary statistics for approximate {B}ayesian
  computation: semi-automatic approximate {B}ayesian computation (with
  discussion).
\newblock \emph{Journal of the Royal Statistical Society series B},
  74:\penalty0 419--474, 2012.

\bibitem[Forman and S{\o}rensen(2014)]{fs:14}
J.~L. Forman and M.~S{\o}rensen.
\newblock A transformation approach to modelling multi-modal diffusions.
\newblock \emph{Journal of Statistical Planning and Inference}, 146:\penalty0
  56--69, 2014.

\bibitem[Girolami et~al.(2013)Girolami, Lyne, Strathmann, Simpson, and
  Atchade]{girolami2013playing}
M.~Girolami, A.~M. Lyne, H.~Strathmann, D.~Simpson, and Y.~Atchade.
\newblock Playing {R}ussian roulette with intractable likelihoods.
\newblock 2013.
\newblock \texttt{arXiv:1306.4032}.

\bibitem[Golightly and Wilkinson(2011)]{golightly2011bayesian}
A.~Golightly and D.~J. Wilkinson.
\newblock Bayesian parameter inference for stochastic biochemical network
  models using particle {M}arkov chain {M}onte {C}arlo.
\newblock \emph{Interface Focus}, 1\penalty0 (6):\penalty0 807--820, 2011.

\bibitem[Gordon et~al.(1993)Gordon, Salmond, and Smith]{gordon1993novel}
N.~J. Gordon, D.~J. Salmond, and A.~F.~M. Smith.
\newblock Novel approach to nonlinear/non-{G}aussian {B}ayesian state
  estimation.
\newblock \emph{IEE PROCEEDINGS-F}, 140\penalty0 (2):\penalty0 107--113, 1993.

\bibitem[Haario et~al.(2001)Haario, Saksman, and Tamminen]{haario-et-al(2001)}
H.~Haario, E.~Saksman, and J.~Tamminen.
\newblock An adaptive {M}etropolis algorithm.
\newblock \emph{Bernoulli}, 7\penalty0 (2):\penalty0 223--242, 2001.

\bibitem[Ionides et~al.(2006)Ionides, Bret{\'o}, and
  King]{ionides2006inference}
E.~L. Ionides, C.~Bret{\'o}, and A.~A. King.
\newblock Inference for nonlinear dynamical systems.
\newblock \emph{Proceedings of the National Academy of Sciences}, 103\penalty0
  (49):\penalty0 18438--18443, 2006.

\bibitem[Kloeden and Platen(1992)]{kloeden-platen(1992)}
P.~E. Kloeden and E.~Platen.
\newblock \emph{Numerical Solution of Stochastic Differential Equations}.
\newblock Springer, 1992.

\bibitem[Korattikara et~al.(2014)Korattikara, Chen, and
  Welling]{korattikara2013austerity}
A.~Korattikara, Y.~Chen, and M.~Welling.
\newblock Austerity in {MCMC} land: cutting the {M}etropolis-{H}astings budget.
\newblock 2014.
\newblock \texttt{arXiv:1304.5299}.

\bibitem[Lee and Andrieu(2012)]{lee-andrieu}
A.~Lee and C.~Andrieu.
\newblock Discussion of ``{C}onstructing summary statistics for approximate
  {B}ayesian computation: semi-automatic approximate {B}ayesian computation''.
\newblock \emph{Journal of the Royal Statistical Society series B},
  74:\penalty0 419--474, 2012.

\bibitem[Lenormand et~al.(2013)Lenormand, Jabot, and
  Deffuant]{lenormand2013adaptive}
M.~Lenormand, F.~Jabot, and G.~Deffuant.
\newblock Adaptive approximate {B}ayesian computation for complex models.
\newblock \emph{Computational Statistics}, 28\penalty0 (6):\penalty0
  2777--2796, 2013.

\bibitem[Marin et~al.(2012)Marin, Pudlo, Robert, and Ryder]{marin-et-al(2011)}
J.~M. Marin, P.~Pudlo, C.~P. Robert, and R.~Ryder.
\newblock Approximate {B}ayesian computational methods.
\newblock \emph{Statistics and Computing}, 22\penalty0 (6):\penalty0
  1167--1180, 2012.

\bibitem[Murray(2013)]{murray2013bayesian}
L.M. Murray.
\newblock Bayesian state-space modelling on high-performance hardware using
  {LibBi}.
\newblock \texttt{arXiv:1306.3277}, 2013.

\bibitem[Newey(1990)]{newey1990semiparametric}
W.K. Newey.
\newblock Semiparametric efficiency bounds.
\newblock \emph{Journal of Applied Econometrics}, 5\penalty0 (2):\penalty0
  99--135, 1990.

\bibitem[Pavliotis and Stuart(2007)]{PavStuart}
G.~A. Pavliotis and A.~M. Stuart.
\newblock Parameter estimation for multiscale diffusions.
\newblock \emph{Journal of Statistical Physics}, 127:\penalty0 741--781, 2007.

\bibitem[Picchini(2013)]{abc-sde}
U.~Picchini.
\newblock \texttt{abc-sde}: a \textsc{Matlab} toolbox for approximate
  {B}ayesian computation ({ABC}) in stochastic differential equation models,
  2013.
\newblock \url{http://sourceforge.net/projects/abc-sde/}.

\bibitem[Picchini(2014)]{picchini(2012)}
U.~Picchini.
\newblock Inference for {SDE} models via approximate {B}ayesian computation.
\newblock \emph{Journal of Computational and Graphical Statistics}, 23\penalty0
  (4):\penalty0 1080--1100, 2014.

\bibitem[Pokern et~al.(2009)Pokern, Stuart, and Wiberg]{stuart}
Y.~Pokern, A.~M. Stuart, and P.~Wiberg.
\newblock Parameter estimation for partially observed hypoelliptic diffusions.
\newblock \emph{Journal of the Royal Statistical Society series B},
  71:\penalty0 49--73, 2009.

\bibitem[Pritchard et~al.(1999)Pritchard, Seielstad, Perez-Lezaun, and
  Feldman]{pritchard1999population}
J.~K. Pritchard, M.~T. Seielstad, A.~Perez-Lezaun, and M.~W. Feldman.
\newblock Population growth of human {Y} chromosomes: a study of {Y} chromosome
  microsatellites.
\newblock \emph{Molecular Biology and Evolution}, 16\penalty0 (12):\penalty0
  1791--1798, 1999.

\bibitem[R{\"o}{\ss}ler(2010)]{rossler(2010)}
A.~R{\"o}{\ss}ler.
\newblock Runge-{K}utta methods for the strong approximation of solutions of
  stochastic differential equations.
\newblock \emph{SIAM Journal on Numerical Analysis}, 48\penalty0 (3):\penalty0
  922--952, 2010.

\bibitem[Socci et~al.(1996)Socci, Onuchic, and Wolynes]{sow:96}
N.D. Socci, J.~N. Onuchic, and P.~G. Wolynes.
\newblock Diffusive dynamics of the reaction coordinate for protein folding
  funnels.
\newblock \emph{Journal of Chemical Physics}, 104:\penalty0 5860--5868, 1996.

\bibitem[Toni et~al.(2009)Toni, Welch, Strelkowa, Ipsen, and
  Stumpf]{toni2009approximate}
T.~Toni, D.~Welch, N.~Strelkowa, A.~Ipsen, and M.~P.~H. Stumpf.
\newblock Approximate {B}ayesian computation scheme for parameter inference and
  model selection in dynamical systems.
\newblock \emph{Journal of the Royal Society Interface}, 6\penalty0
  (31):\penalty0 187--202, 2009.

\bibitem[Varin et~al.(2011)Varin, Reid, and Firth]{varin2011overview}
C.~Varin, N.~Reid, and D.~Firth.
\newblock An overview of composite likelihood methods.
\newblock \emph{Statistica Sinica}, 21\penalty0 (1):\penalty0 5--42, 2011.

\bibitem[Wilkinson(2012)]{d.wilkinson(2012)}
D.~J. Wilkinson.
\newblock \emph{Stochastic Modelling for Systems Biology}.
\newblock CRC Press, second edition, 2012.

\bibitem[Wolynes et~al.(2012)Wolynes, Eaton, and Fersht]{wolynes2012chemical}
P.~G. Wolynes, W.~A. Eaton, and A.~R. Fersht.
\newblock Chemical physics of protein folding.
\newblock \emph{Proceedings of the National Academy of Sciences}, 109\penalty0
  (44):\penalty0 17770--17771, 2012.

\end{thebibliography}
\bibliographystyle{plainnat}

\end{document}